\documentclass[%
 aip,
 amsmath,amssymb,
 reprint,%
]{revtex4-2}
\usepackage{amsmath}
\usepackage{subcaption}
\usepackage{hyperref}
\usepackage{bm}
\usepackage{graphicx}
\usepackage{dcolumn}
\usepackage{bm}
\usepackage[utf8]{inputenc}
\usepackage[T1]{fontenc}
\usepackage{mathptmx}
\usepackage{etoolbox}
\usepackage{csvsimple}
\usepackage{microtype}
\usepackage[normalem]{ulem}
 \useunder{\uline}{\ul}{}

 \usepackage{array, caption, tabularx,  ragged2e,  booktabs}
 \usepackage{mathtools}
\newcommand{\lgradb}{{$L_{ \nabla \mathbf{B}}$}}

\usepackage{cleveref}
\usepackage{tikz}
\usetikzlibrary{shapes.geometric, arrows}

\crefname{equation}{}{}
\creflabelformat{equation}{(#2#1#3)}

\tikzstyle{start} = [rectangle, rounded corners, 
minimum width=1cm, 
minimum height=1cm,
text centered, 
draw=black, 
text width=35mm,
fill=white!30]

\tikzstyle{stop} = [rectangle, rounded corners, 
minimum width=1cm, 
minimum height=1cm,
text centered, 
draw=black, 
fill=white!30]

\tikzstyle{io} = [rectangle, 
minimum height=1cm,
text centered,
text width=3cm,
draw=black, fill=white!30]

\tikzstyle{process} = [rectangle, 
minimum width=3cm, 
minimum height=1cm, 
text centered, 
text width=3cm, 
draw=black, 
fill=white!30]

\tikzstyle{connect} = [circle, 
text centered, 
draw=black, 
fill=white!30]

\tikzstyle{decision} = [diamond,
aspect = 1.5,
minimum height=2cm, 
text centered, 
draw=black, 
fill=white!30]
\tikzstyle{arrow} = [thick,->,>=stealth]

\makeatletter
\def\@email#1#2{%
 \endgroup
 \patchcmd{\titleblock@produce}
  {\frontmatter@RRAPformat}
  {\frontmatter@RRAPformat{\produce@RRAP{*#1\href{mailto:#2}{#2}}}\frontmatter@RRAPformat}
  {}{}
}%
\makeatother
\begin{document}

\preprint{AIP/123-QED}

\title[Magnetic Gradient Scale Length]{The Magnetic Gradient Scale Length Explains Why Certain Plasmas Require Close External Magnetic Coils}
\author{John Kappel}
\author{Matt Landreman}%
\email{Jkappel@umd.edu}
\affiliation{ 
Institute for Research in Electronics and Applied Physics, University of Maryland, College Park, MD
20742, United States of America
}%

\author{Dhairya Malhotra}
\affiliation{Flatiron Instituite, New York, NY, 10010, United States of America
}%

\date{\today}

\begin{abstract}
The separation between the last closed flux surface of a plasma and the external coils that magnetically confine it is a limiting factor in the construction of fusion-capable plasma devices. This plasma-coil separation must be large enough so that components such as a breeding blanket and neutron shielding can fit between the plasma and the coils. Plasma-coil separation affects reactor size, engineering complexity, and particle loss due to field ripple. For some plasmas it can be difficult to produce the desired flux surface shaping with distant coils, and for other plasmas it is infeasible altogether. Here, we seek to understand the underlying physics that limits plasma-coil separation and explain why some configurations require close external coils. In this paper, we explore the hypothesis that the limiting plasma-coil separation is set by the shortest scale length of the magnetic field as expressed by the $\nabla \mathbf{B}$ tensor. We tested this hypothesis on a database of $>$ 40 stellarator and tokamak configurations. Within this database, the coil-to-plasma distance compared to the minor radius  varies by over an order of magnitude. The magnetic scale length is well correlated to the coil-to-plasma distance of actual coil designs generated using the \texttt{REGCOIL} method [Landreman, Nucl. Fusion 57, 046003 (2017)]. Additionally, this correlation reveals a general trend that larger plasma-coil separation is possible with a small number of field periods.
 \end{abstract}

\maketitle

\section{Introduction}
\label{sec:Introduction}

Magnetic fields are used to confine plasmas at the necessary temperature and density for fusion. The external coils that generate part of these magnetic fields are subject to engineering constraints in order for the device to be feasible to build. One engineering constraint relates to the separation between the last closed flux surface of the plasma and the edge of the nearest external magnetic coil. In reactors, neutron shielding is needed between the plasma and the coils. This shielding protects the magnets from damage caused by the high energy neutrons released from fusion reactions. In addition, due to the limited tritium available worldwide, a reactor requires a breeding blanket to produce more tritium. These components require a separation of roughly 1.5 m between the last closed flux surface (LCFS) and the nearest coil, so that the there is room for both the neutron shielding and the breeding blanket.\cite{ARIES-CS_report} In this paper, we shall refer to this distance as plasma-coil separation.

A larger plasma-coil separation also desirable since insufficient space can lead to increased costs during  design and construction. \cite{W7-X_report} Even in experiments without neutron shielding or a breeding blanket, miscellaneous  components are installed between the plasma and the magnets with little separation in between each component. When the vacuum chamber's pressure is lowered, when the magnets are cooled to superconducting temperature, or when the magnets are energized, these components can shift and potentially collide. Additional plasma-coil clearance can allow for components to be spaced further apart, simplifying the engineering design and assembly.

\begin{figure*}
     \centering
     \begin{subfigure}{0.30\textwidth}
         \centering
         \includegraphics[width=\textwidth]{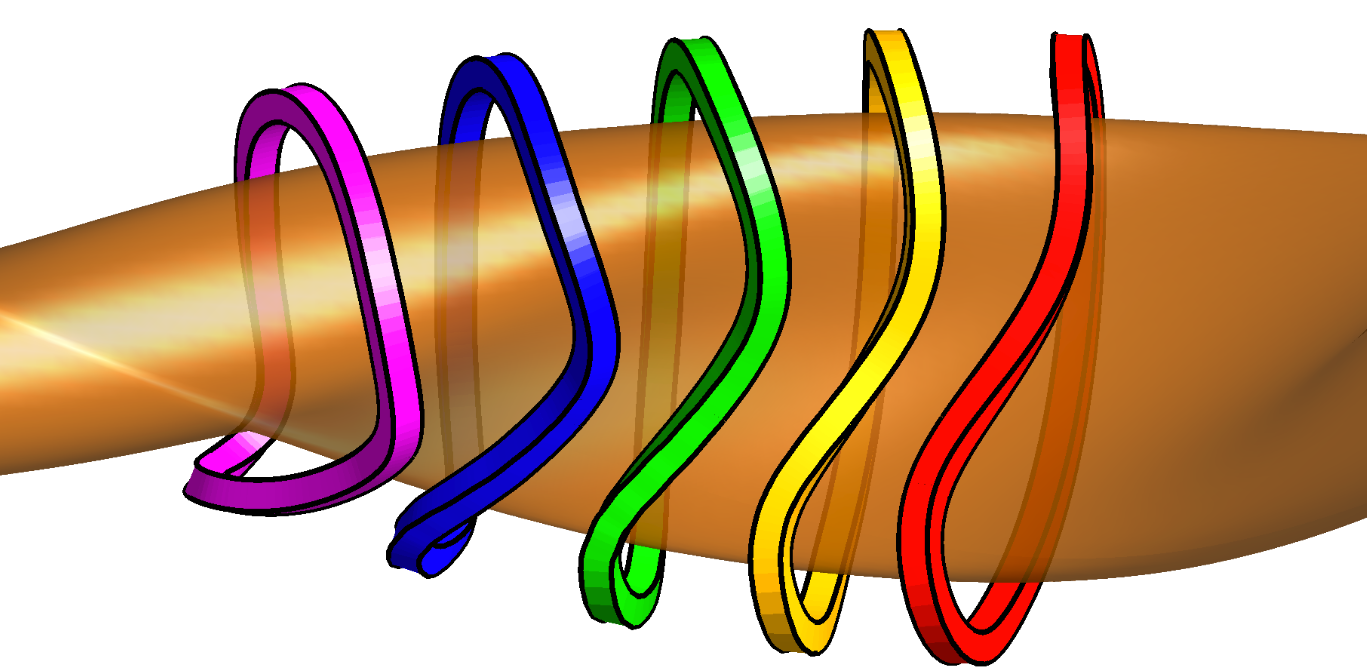}
         \caption{}
     \end{subfigure}
    \hspace{4 mm}
     \begin{subfigure}{0.31\textwidth}
         \centering
         \includegraphics[width=\textwidth]{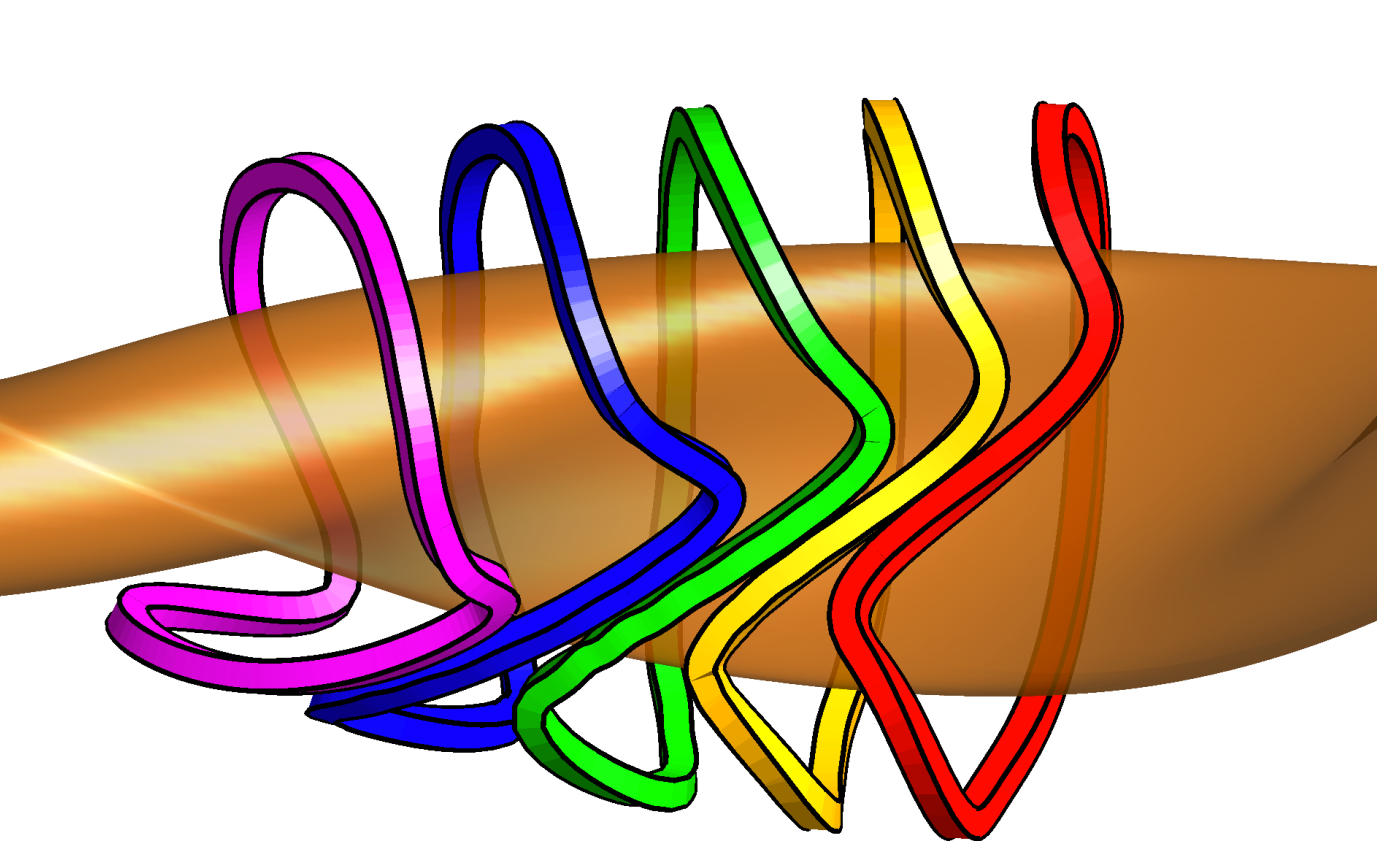}
         \caption{}
     \end{subfigure}
         \hspace{4 mm}
        \begin{subfigure}{0.32\textwidth}
         \centering
         \includegraphics[width=\textwidth]{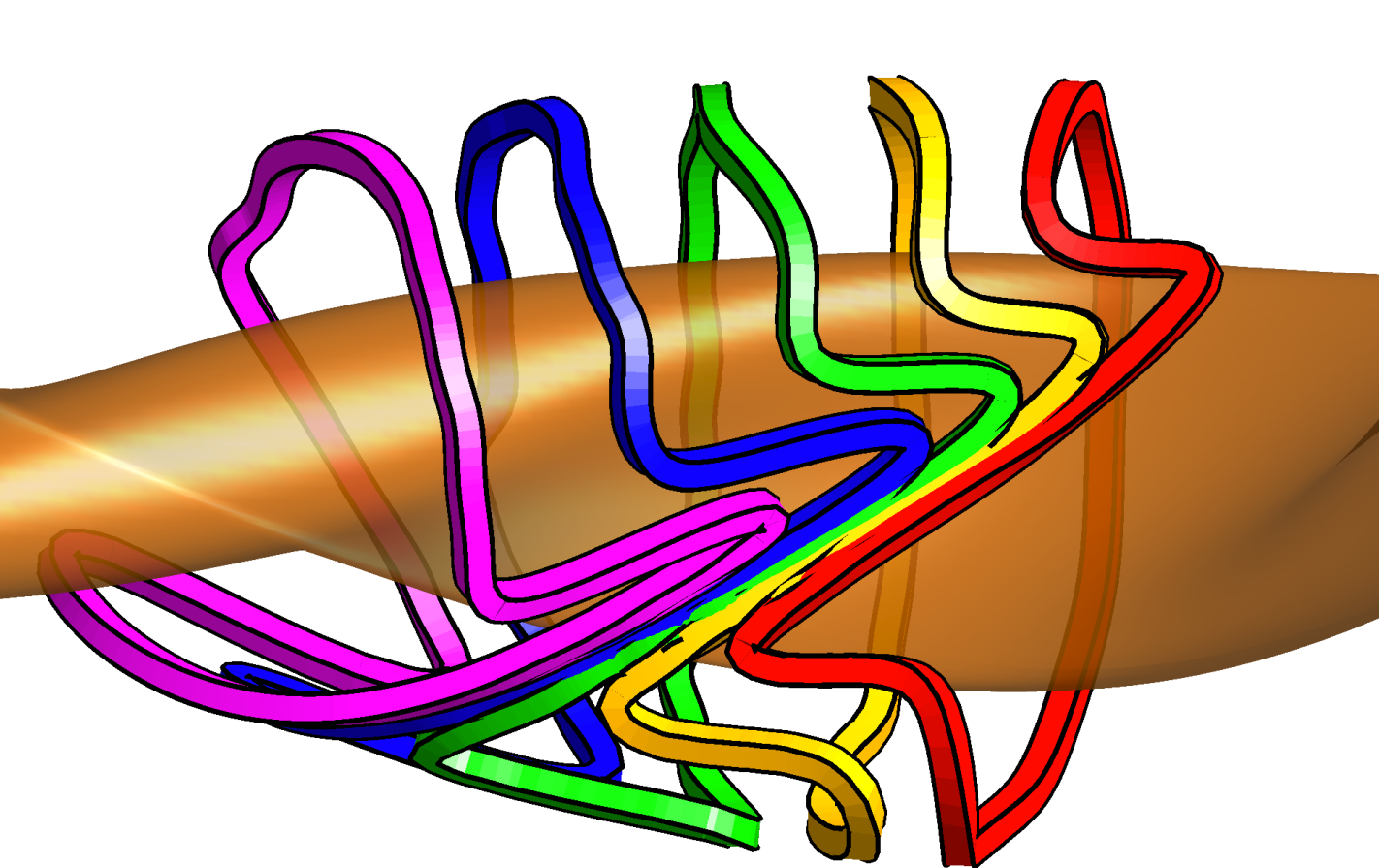}
         \caption{}
          \end{subfigure}
     \hfill
     \caption{Using \texttt{REGCOIL}, coils are calculated at constant distances of (a) 25 cm, (b) 50 cm, and (c) 65 cm away from the last closed flux surface to create the magnetic field for the W7-X plasma (which has a minor radius of $a$ = 53 cm). The generated flux surfaces all have a fixed $B_{\mathrm{RMS}}$ error of 0.02 T. At higher offset distances, the coils have increased complexity.}
     \label{fig:CoilComplexity}
\end{figure*}

Increasing the minimum plasma-coil distance has additional advantages. Low plasma-coil separation causes ripple in the field strength, which can lower confinement.\cite{10.1063/1.4876740} In addition, because coils must be built at a minimum clearance from the plasma, achieving a larger plasma-coil separation can counter-intuitively allow for smaller reactors and hence reduced costs. One can scale down a plant design with a large plasma-coil separation until the separation is around 1.5 m. Thus, in several previous reactor studies, coil-plasma separation determines the minimum reactor size. \cite{ARIES-CS_report, Lion_2021}

 However, it can be difficult to optimize coils to achieve a large plasma-coil separation, especially in stellarators. Stellarator optimization is typically divided into two stages. In stage I, the boundary shape of the plasma is optimized in order to achieve properties such as confinement and stability. In stage II,  the coil shapes are optimized in order to generate the target field designed in stage I, subject to engineering constraints.  Due to the complex shape of a stellarator plasma, moving coils farther away during stage II optimization at a fixed field accuracy results in increased coil complexity (such as increased curvature, longer coils, and closer minimum coil-coil distance), as shown in figure \ref{fig:CoilComplexity}. This concept is further elaborated later in this paper in figure \ref{fig:KMaxScan}. Conversely, increasing plasma-coil separation at a fixed coil complexity results in coils that do not reproduce the target plasma shape to a high enough level of accuracy. A partial solution to this issue may be to perform single-stage optimization, in which both the plasma shape and coil shapes are optimized together.\cite{Drevlak_2019,GIULIANI2022111147,Jorge_2023,henneberg_hudson_pfefferlé_helander_2021} However, single-stage optimization can be computationally challenging. Single-stage optimization also strongly benefits from a good initial condition. It remains valuable therefore to develop an easy-to-calculate proxy for plasma-coil separation for a given plasma shape before a detailed coil design is found.

In this paper, we explore the hypothesis that differences in plasma-coil separation between magnetic configurations can be understood in terms of scale lengths of the magnetic field. We demonstrate this relationship by comparing the magnetic gradient scale length to the plasma-coil separation for a diverse database of plasma configurations, and we show a good correlation between the two. For the coil calculations, we use the  \texttt{REGCOIL} method,\cite{Landreman_2017,REGCOIL} which is convenient for performing a systematic comparison of many configurations. Because of this good correlation, we propose that the smallest magnetic gradient scale length on the surface of the plasma can be used as a proxy for the minimum plasma-coil separation.
If a configuration has a small magnetic gradient scale length, it will be intrinsically difficult to find distant coils no matter what coil design code is used.

 In section \ref{section:Construction}, we motivate a quantitative measure of the magnetic gradient scale length, and consider this magnetic gradient scale length in model geometry. In section \ref{section:Methods}, we explain the methods used to calculate the coil configurations for the database of $\sim$ 40 plasma configurations in order to observe the agreement between the calculated plasma-coil separation and the magnetic gradient scale length. In section \ref{section:Results}, we demonstrate the excellent agreement between the two lengths, and provide a discussion of the results. Finally, we conclude in section \ref{Sec:Summary}.

\section{Motivation for the \lgradb~Scale Length}
\label{section:Construction}

Scale lengths are a common tool in plasma physics. When deriving models, we often employ arguments involving scale lengths to determine which effects are negligible and which effects are significant. For example, when deriving the gyrokinetic or drift-kinetic equations, terms are ordered in the parameter $\rho/L$, where $\rho$ is the gyroradius and $L$ is a macroscopic scale length of the equilibrium, $L \sim n/ |\nabla n| \sim T / | \nabla T| \sim B/|\nabla B|$.\cite{Howes_2006} For any scalar $Q$, it is possible to write a gradient scale length of the form
\begin{equation}
     \frac{Q}{ \| \nabla Q\|}.
\end{equation}
One can attempt to design a scale length of similar form for the magnetic field. However, the magnetic field is not a scalar but rather a vector, and so the most useful expression for the magnetic gradient scale length is not immediately obvious.

One quantity resembling a scale length for magnetic fields was discussed by Nara et al. and in subsequent papers.\cite{Nara,10.1117/1.JRS.8.083596,doi:10.1190/1.2164759,7366564,10.1190/1.1442807} In these papers, an equation was derived for the location of a dipole source, based on the local field and its gradient:
\begin{equation}
\mathbf{r} = -3 (\nabla \mathbf{B})^{-1} \mathbf{B},
\label{naraeq}
\end{equation}
where the tensor $\nabla \mathbf{B}$ can be expressed as a $3\times 3$ matrix:
\begin{equation}
    \nabla \mathbf{B} =\begin{bmatrix}
\frac{\partial B_x}{\partial x} & \frac{\partial B_y}{\partial x} & \frac{\partial B_z}{\partial x}\\
\frac{\partial B_x}{\partial y} & \frac{\partial B_y}{\partial y} & \frac{\partial B_z}{\partial y} \\
\frac{\partial B_x}{\partial z} & \frac{\partial B_y}{\partial z} & \frac{\partial B_z}{\partial z}
\end{bmatrix}.
\label{gradbmatrix}
\end{equation}
 Equation (\ref{naraeq}) can be solved to localize a magnetic source, and has applications in the field of RFID tag positioning and motion tracking.\cite{DipolePaper} While this equation does further prove that the gradient of the magnetic field vector can be used for source localization, this equation assumes a single dipole-like source. Moreover, the matrix $\nabla \mathbf{B}$ is not invertible for a straight wire, meaning (\ref{naraeq}) is a poor choice for an arrangement of coils. To find a more suitable magnetic gradient scale length, let us consider some desirable properties of it.

We first would like all components of the $\nabla\mathbf{B}$ matrix to significantly contribute to the scale length. As an example without this property, consider the possible scale length candidate
\begin{equation}
 \frac{B}{\| \nabla B \|}.
 \label{eq:scalarB}
\end{equation}
Equation (\ref{eq:scalarB}) can be written explicitly in terms of Cartesian components of the magnetic field vector and its gradient matrix:
\begin{equation}
    \begin{split}
        \frac{B}{\| \nabla B \|} = B^2 \biggl[ \biggl(B_x \frac{\partial B_x}{\partial x} + B_y \frac{\partial B_y}{\partial x} + B_z \frac{\partial B_z}{\partial x}  \biggr)^{2}
        \\+ \biggl(B_x \frac{\partial B_x}{\partial y} + B_y \frac{\partial B_y}{\partial y} + B_z \frac{\partial B_z}{\partial y}\biggr]^{2} \\
        + \biggl(B_x \frac{\partial B_x}{\partial z} + B_y \frac{\partial B_y}{\partial z} + B_z \frac{\partial B_z}{\partial z}\biggr)^{2}\biggr)^{-1/2}.
    \end{split}  
     \label{eq:scalarExpanded}
\end{equation}
Consider a point $(x_1,y_1,z_1)$ where the magnetic field is \textit{locally} oriented in a single coordinate direction, i.e. $\mathbf{B}(x_1,y_1,z_1)= \hat{\mathbf{z}} B_z(x_1,y_1,z_1)$. Because $B_x(x_1,y_1,z_1,) = B_y(x_1,y_1,z_1,) = 0$, equation (\ref{eq:scalarExpanded}) can be simplified to
\begin{equation}
\frac{B}{\| \nabla B \|} =
    \frac{B^2}{B_z\biggl(\frac{\partial B_z}{\partial x}^2 + \frac{\partial B_z}{\partial y}^2 + \frac{\partial B_z}{\partial z}^2\biggr)^{1/2}}.
\end{equation}
Only three of the nine elements of the $\nabla\mathbf{B}$ matrix appear.
Consider another point $(x_2,y_2,z_2)$ where the magnetic field differs from point $(x_1,y_1,z_1)$ only in the gradient, i.e.:
\begin{equation}
    \frac{\partial B_x}{\partial x}\bigg\rvert_{(x_1,y_1,z_1)} \neq \frac{\partial B_x}{\partial x}\bigg\rvert_{(x_2,y_2,z_2)}.
\end{equation}
Although the gradients of their magnetic vectors differ, the scale length proposed in equation (\ref{eq:scalarB}) would be the same at both points. In this example, six of the nine Cartesian components of the gradient matrix would have no impact on the scale length. This behavior makes equation (\ref{eq:scalarB}) a poor choice of a scale length.
The same consideration rules out another plausible scale length, the radius of curvature of the magnetic field:
\begin{equation}
\frac{1}{\| \mathbf{b} \cdot  \nabla \mathbf{b} \|},
\label{eq: radius of curvature}
\end{equation}
where $\mathbf{b} = \mathbf{B}/B$ is the unit vector in the direction of the field.\cite{d1991flux} A similar argument can be used to prove that this scale length is independent of some Cartesian components of $\nabla\mathbf{B}$ when $\mathbf{B}$ is aligned with a coordinate axis. 
Therefore, we are motivated to instead base our scale length on a matrix norm of the gradient matrix $\nabla \mathbf{B}$.

\begin{figure*}
     \centering
     \begin{subfigure}{0.24\textwidth}
         \centering
         \includegraphics[width=\textwidth]{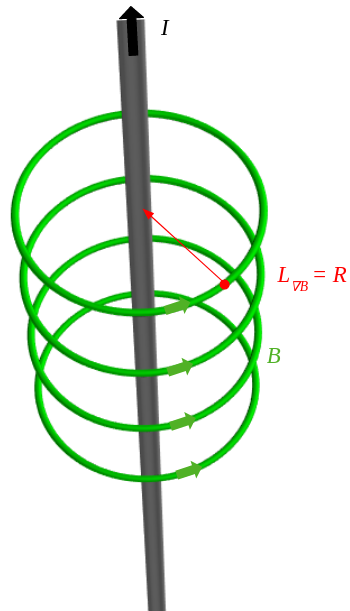}
         \caption{}
         \label{fig:wirelabled}
     \end{subfigure}
    \hspace{10 mm}
     \begin{subfigure}{0.57\textwidth}
         \centering
         \includegraphics[width=\textwidth]{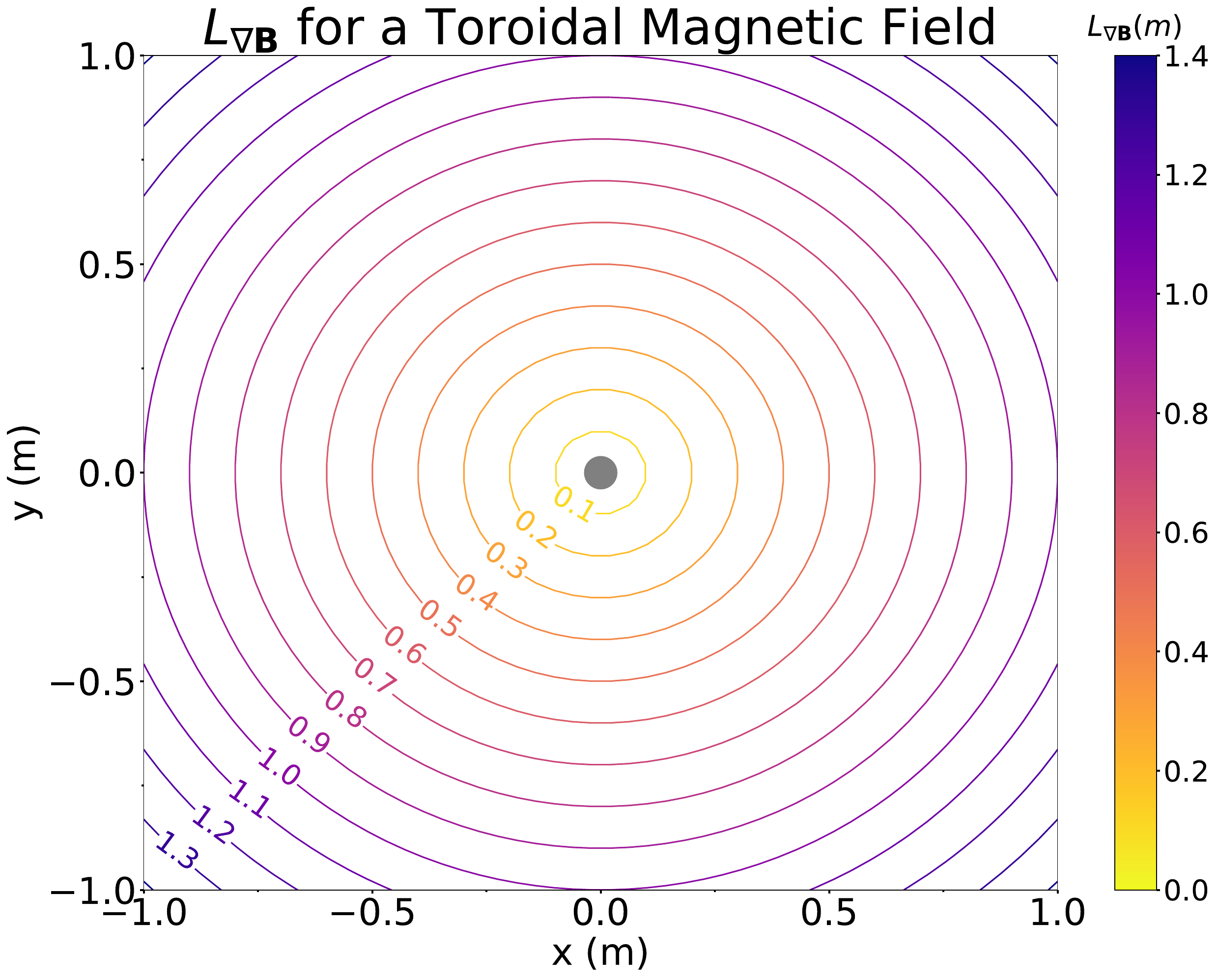}
         \caption{}
         \label{fig:toroidalfield}
     \end{subfigure}
     \hfill
     \caption{(a) A For an infinite straight wire, the \lgradb~scale length calculated from the local magnetic field (for example the red dot) is the distance to the wire. One can compute \lgradb~without knowing the location of the wire. (b) At any point around the wire (represented by the gray dot), we can use \lgradb~to find the distance to the wire by only measuring $B$ and $ \nabla \mathbf{B}$ at one point.}
     \label{fig:straightwire}
\end{figure*}

Our second consideration is that the norm should be invariant to rotation of coordinates. In other words, given some unitary rotation matrix $\mathsf{U}$, $ \| \mathsf{U} \nabla \mathbf{B} \| = \| \nabla \mathbf{B} \|$. Norms that fulfill this property are referred to as unitarily invariant norms. Two common classes of unitarily invariant norms are Schatten\cite{Schatten} and Ky Fan\cite{fan1949theorem} norms. For a matrix $\mathsf{A}$ with dimensions $m \times n$, the Schatten-$p$ norm is defined as
\begin{equation}
    \|\mathsf{A}\|^S_p =\biggl ( \sum_{i=1}^n  \sigma_i^p \biggr)^{1/p},
\end{equation}
and the Ky Fan-$k$ norm is defined as
\begin{equation}
    \|\mathsf{A}\|^K_k = \sum_{i=1}^k  \sigma_i,
\end{equation}
where $\sigma_i$ is the $i$-th singular value of $\mathsf{A}$, $p \geq 1$, and $1 \leq k \leq n$. The most common unitarily invariant norm is Frobenius norm:
\begin{equation}
\|\mathsf{A}\|_F =\sqrt{\sum_{i=1}^m \sum_{j=1}^n |a_{ij}|^2},
\end{equation}
where $a_{ij}$ are the elements of $\mathsf{A}$. 
Note that for the gradient of a magnetic field, the Frobenius norm is equivalent to the Schatten-2 norm in a vacuum field. In a vacuum field, the magnetic field $\mathbf{B}$ is curl-free and so $\mathbf{B} = \nabla \Phi$ for some scalar potential $\Phi$. Therefore,$\nabla \mathbf{B} = \nabla \nabla \Phi$, which is manifestly symmetric.
While the Frobenius norm is not the only unitarily invariant norm, we will adopt it in this work for simplicity.

Thus, given our two criteria for the magnetic scale length -- a significant contribution from each component of the $\nabla \mathbf{B}$ matrix and rotational invariance -- we are led to the following formula for the magnetic gradient scale length:
\begin{equation}
    L_{\nabla \mathbf{B}}=  \frac{\sqrt{2} B}{  \| \nabla \mathbf{B} \|_F}.
    \label{eq:lgradb}
\end{equation}
 The $\sqrt{2}$ factor is explained below.
An alternative candidate scale length is 
 \begin{equation}
      L_{\|\sigma\|} = \frac{\sqrt{2} B}{ ( \Sigma_i \sigma_i^2)^{1/2}},
      \label{eq:lsingularb}
 \end{equation}
 where $\sigma_i$ are the 3 singular values of the matrix $\nabla \mathbf{B}$. However, in a current-free field, $\nabla \mathbf{B}$ is symmetric, so equations \ref{eq:lgradb} and \ref{eq:lsingularb} are equivalent.  It is worth noting that equation (\ref{eq:lgradb}) has proven useful for optimizing near axis configurations. \cite{landreman_2021,landreman_2022,jorge_2022} In Appendix \ref{sec:scalelength}, we compare equation (\ref{eq:lgradb}) to alternative possible measures of a magnetic gradient scale length.

To develop a geometric intuition of $L_{\nabla \mathbf{B}},$ let us look at the example of a  magnetic field generated by an infinite straight wire. The magnetic field in cylindrical coordinates ($ R, \phi, z$) is
\begin{equation}
  \mathbf{B}( R) = \frac{\mu_0 I}{2 \pi R}  \hat{\pmb \phi},  
\end{equation}
where $I$ is the current of the wire, and the hat symbols indicate unit vectors.  Because of its simplicity, the model geometry of straight wire is a good initial test of the effectiveness of the \lgradb~scale length.
Let us determine the scale length analytically: 
\begin{gather}
 \label{eq: wire1}
   \nabla \mathbf{B} = - \frac{\mu_0 I} {2 \pi R^2} ( \hat{\pmb \phi} \hat{\mathbf{R}} + \hat{\mathbf{R}} \hat{\pmb \phi}),\\
    \label{eq: wire2}
   \|\nabla \mathbf{B} \|_F = \frac{\sqrt{2} \mu_0 I}{2 \pi R^2},\\
    \label{eq: wire3}
   L_{\nabla \mathbf{B}}= \sqrt{2} \frac{B}{\| \nabla \mathbf{B} \|_F} = R.
\end{gather}
Therefore, at any point in the magnetic field, \lgradb~is equal to the distance away from the straight wire. This relationship is shown in figure \ref{fig:straightwire}. The factor of $\sqrt{2}$ was included in equation (\ref{eq:lgradb}) to rescale \lgradb~so that it is equivalent with the distance to a wire in this reduced model. Note that equation \cref{eq: wire3} does not require knowing where the coil is located, only local measurement of $B$ and $\nabla \mathbf{B}$. This property means that we can use these two quantities to approximate the distance to the nearest coil. See Appendix \ref{sec:circular} for a discussion of \lgradb~for the magnetic field of a circular wire.

\section{Methods of Numerical Validation}
\subsection{Requirements for Unique Coil Optimization}

\label{section:Methods}

In order to verify the accuracy of \lgradb~as a proxy for the plasma-coil separation, we compare \lgradb~to the plasma-coil separation of stage II optimized stellarator configurations. However, the coils that generate a given plasma shape are not unique. Given a magnetic field $\mathbf{B(x)}$ which is only known within a region that is a finite distance away from the current-carrying source, for any $\delta > 0$ there exists an infinite number of current arrangements to generate $\mathbf{B}$ up to an error $\epsilon < \delta$.\cite{Landreman_2017} To ensure a unique coil arrangement, we must establish additional constraints, such as some measure of coil complexity.

There are multiple methods commonly used to compute the shapes of stellarator coils. The most prominent two are filamentary coil methods and current potential methods. In filamentary coil methods, such as the code \texttt{FOCUS},\cite{Zhu_2018}  each coil is parameterized as a curve, allowing for optimization in 3-D space. However, the objective functions of these codes  have multiple local minima, so results are dependent on the initial set of parameters. Also, these methods can have a significant number of adjustable parameters, such as scalarization weights and the number of coils. 
These aspects make it difficult to compute coil shapes in an automated and systematic way for a wide variety of plasma configurations. 
Current potential methods, like \texttt{REGCOIL}, define some preset winding surface on which the coils can lie. These methods optimize the current potential on this surface, and then calculate the coil shapes using contours of the current potential.

 For this paper, we use the  \texttt{REGCOIL} method.\cite{Landreman_2017,REGCOIL}
 \texttt{REGCOIL}, compared to its precursor \texttt{NESCOIL},\cite{Merkel_1987} includes a ridge regression in the current density. Ridge regression not only reduces overfitting, but also ensures the minimum of the loss function in not underdetermined, which guarantees that any local minimum is a unique global minimum. 
\begin{figure*}
     \centering
     \begin{subfigure}{0.33 \textwidth}
         \centering
         \includegraphics[width=1.0\textwidth]{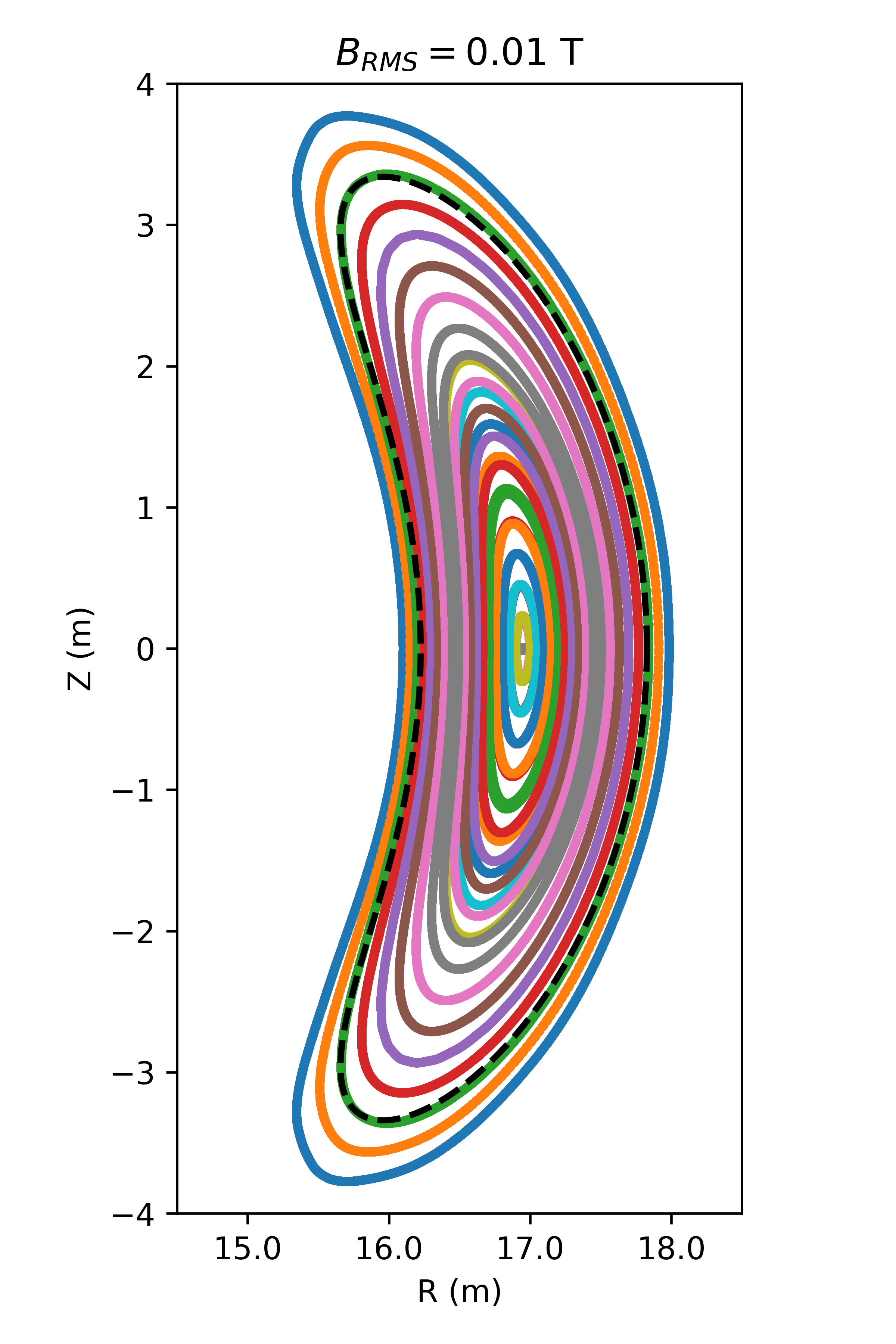}
         \caption{}
         \label{fig: poincare1}
     \end{subfigure}
     \hfill
     \begin{subfigure}{0.33\textwidth}
         \centering
         \includegraphics[width=1.0 \textwidth]{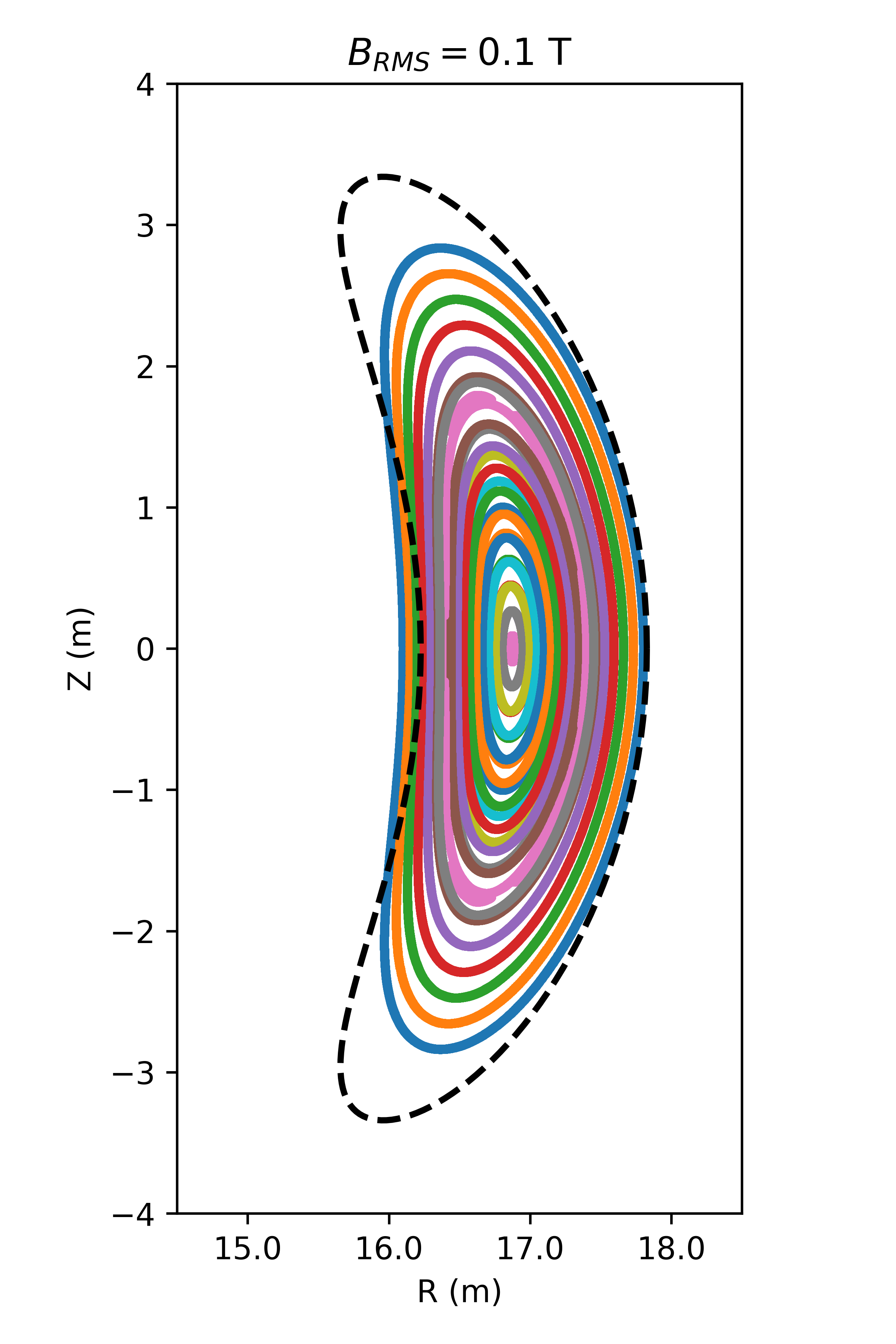}
         \caption{}
         \label{fig: poincare2}
     \end{subfigure}
    \begin{subfigure}{0.33\textwidth}
     \centering
     \includegraphics[width=1.0\textwidth]{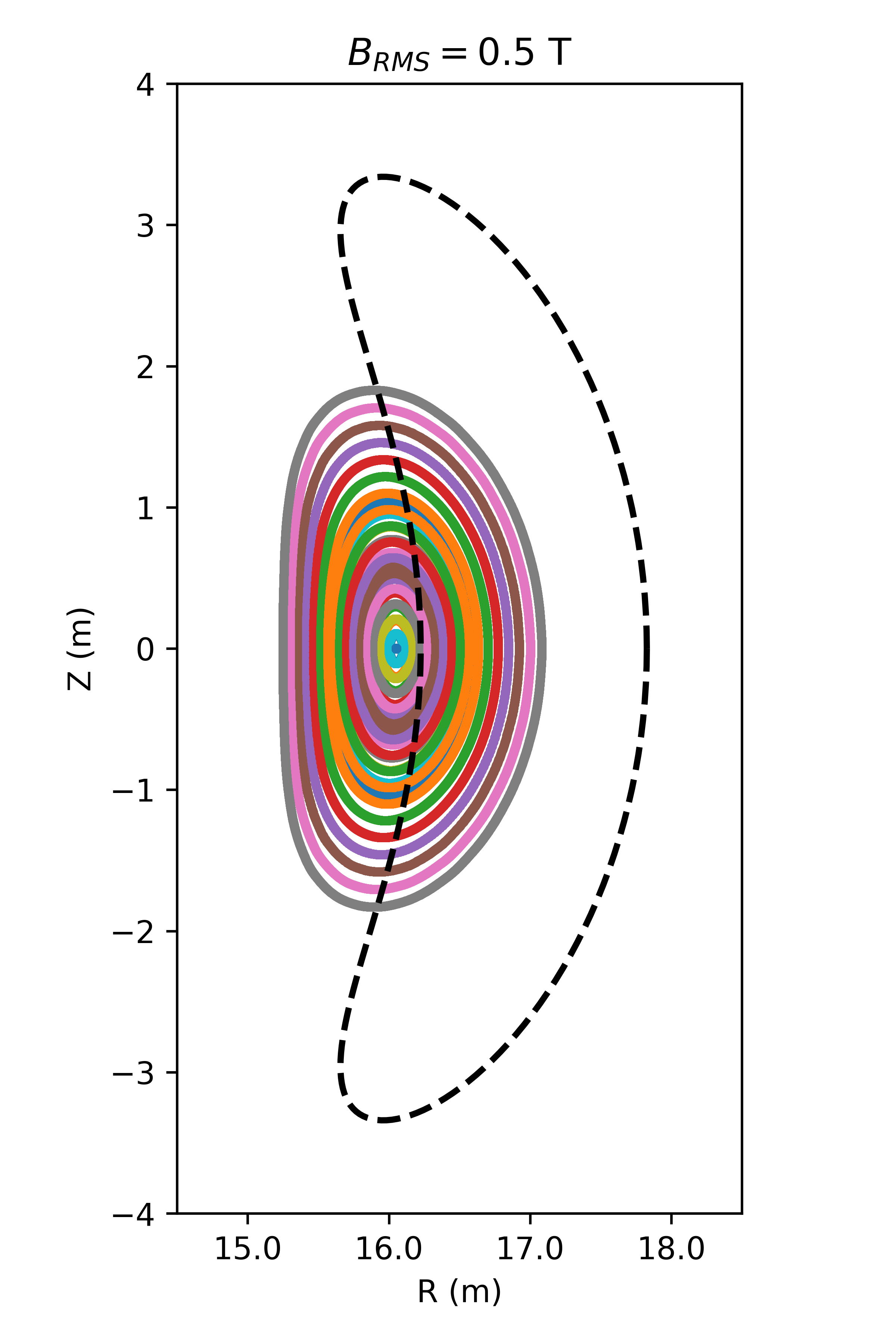}
     \caption{}
     \label{fig: poincare3}
    \end{subfigure}
     \hfill
    \label{fig: poincare}
     \caption{Poincare plots of the precise quasi-helical (QH) stellarator configuration\cite{PhysRevLett.128.035001} scaled to $a$ = 1.704 m and volume averaged $B$ = 5.865 T. The coils for these plots were computed by \texttt{REGCOIL} on a winding surface offset from the plasma surface by 1.796 m. The coils are optimized to achieve a target $B_\mathrm{RMS}$ of (a) 0.01 T, (b) 0.1 T, and (c) 0.5 T. The black dotted line is the target boundary surface.  At high $B_\mathrm{RMS}$, the flux surfaces of the Poincare plot do not match the target flux surface.}
\end{figure*}
Here we present an overview of \texttt{REGCOIL}'s methods.\cite{Landreman_2017} One initializes an infinitely thin winding surface surrounding the plasma. For simplicity, in this paper we define each winding surface to be the surface that is a uniform offset distance $L$ from the plasma surface. One can parameterize the current potential on the winding surface as a Fourier series:
 \begin{equation}
\Phi' (\theta' , \phi') = \sum_j \Phi_j \sin(m_j \theta ' - n_j \phi ') + \frac{G \phi '}{2 \pi} + \frac{I \theta'}{2 \pi},
 \end{equation}
where $\theta'$ and $\phi'$ are the poloidal and toroidal coordinates on the winding surface, $G$ is the total poloidal current, $I$ is the total toroidal current, and assuming stellarator symmetry. For modular coils, $I=0$. From this current potential, we calculate the current density:
\begin{equation}    
  \mathbf{K'} = \mathbf{n'} \times \mathbf{\nabla} \Phi', 
\end{equation}
where $\mathbf{n'}$ is the normal vector on the winding surface. It is then possible to find the magnetic field at any point in space via the Biot-Savart law:
\begin{equation}
\mathbf{B}(\mathbf{r}) = \frac{\mu_0}{4 \pi} \int d^2 a' \frac{ \mathbf{K'} \times \mathbf{(r-r')}}{|\mathbf{r-r}'|^3}.
\end{equation}
\texttt{REGCOIL}'s aim is to find Fourier amplitudes of the potential $\Phi_j$ which minimize the objective function  $f$:
 \begin{equation}
 f = \int d^2a \  (\mathbf{B}(\theta, \phi) \cdot \mathbf{n} )^2 + \lambda \int d^2 a'\  \|\mathbf{K}(\theta ' , \phi ')\|^2, 
 \label{eq:objective}
 \end{equation}
 where $\theta$ and $\phi$ are the poloidal and toroidal coordinates on the plasma surface,  $\mathbf{n}$ is the normal vector on the plasma surface, and $\lambda$ is a regularization parameter. The first term is the quadratic flux over the last closed flux surface. The normal component of the magnetic field, $\mathbf{B\cdot n}$, vanishes throughout a flux surface. Therefore, this term is a measure of disagreement between the magnetic field generated by the external coils and the magnetic field of the target configuration. The second term is a measure of the current density $\mathbf{K}$ over the winding surface, which acts as a $L_2$ regularizer to constrain the otherwise underdetermined system. $\mathbf{K}$ is also a measure of coil complexity, as we discuss further below. Minimizing the objective function will result in coils that generate the desired shape for the LCFS with minimal coil complexity.

\begin{figure}
     \centering
     \includegraphics[width= 0.5 \textwidth]{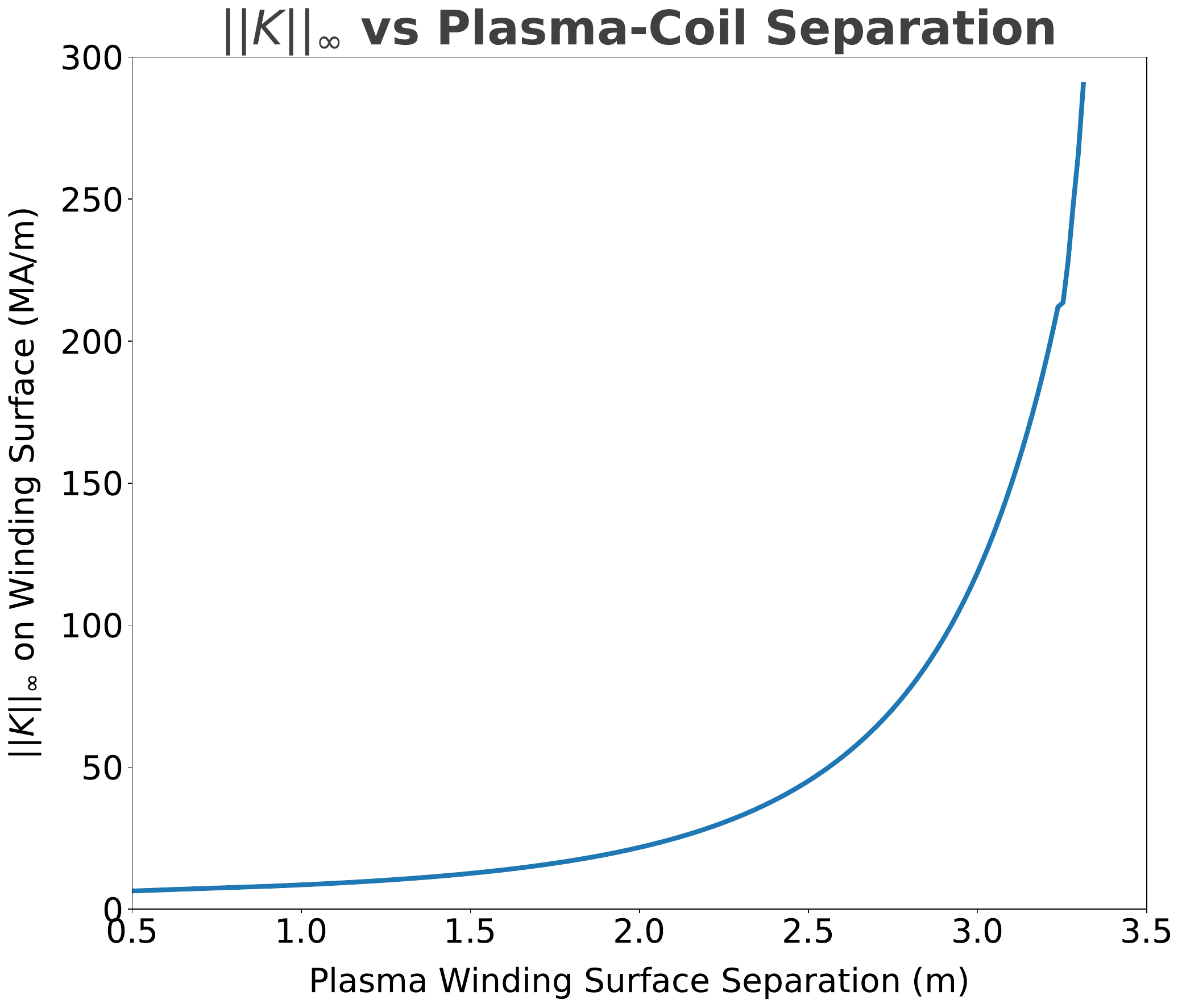}
     \hfill
     \caption{The maximum current density $\|K\|_\infty$ and its dependence on the plasma-coil separation $L$ for the winding surface of a precise QH configuration. For each plasma-coil separation, \texttt{REGCOIL} computes the current potential $\Phi$ on the winding surface that achieves a target $B_\mathrm{RMS}$ = 0.01 T. With this constraint, there is a monotonic relationship between $\|K\|_\infty$ and $L$, so a $\|K\|_\infty$ equality constraint will result in a unique solution for $L$.}
    \label{fig:KMaxScan}
\end{figure}
 
To calculate the current on the winding surface, we need to specify only two parameters, the plasma-coil separation $L$ and the regularization parameter $\lambda$. We choose $L$ and $\lambda$ so that the resulting coils fulfill two constraints. Let us define the quadratic flux normalized to the surface area of the plasma:
\begin{equation}
   B_{\mathrm{RMS}} = \biggl(\frac{\int d^2a \  (\textbf{B} \cdot  \textbf{n})^2 }{A_{\mathrm{plasma}}}\biggr)^{1/2},  
\end{equation}
 where $A_{\mathrm{plasma}}$ is the surface area of the last closed flux surface. Ideally, the normalized quadratic flux is 0 T, but due to the penalty on current density in the objective function, the optimal stage-II magnetic field will not exactly match the stage-I optimized magnetic field. For the results that follow, we chose $\lambda$ so that $B^*_\mathrm{RMS} = 0.01$ T,  with all configurations scaled to an average field strength in the plasma of 5.865 T. The value 0.01 T was chosen as it is low enough to achieve closed flux surfaces that match the target last closed flux surface, as shown in figure \ref{fig: poincare1}.

Let us also define the largest magnitude of the current density $K$ on the winding surface as $\|K\|_{\infty}$. The plasma-coil separation $L$ is adjusted so that $\|K\|_{\infty}$ equals a target value $\|K\|_\infty^*$. This constraint has two useful properties. First, increasing plasma-coil separation while fixing $B^*_{\mathrm{RMS}}$ smoothly increases $\|K\|_{\infty}$ as shown in figure \ref{fig:KMaxScan}. This means the target $\|K\|^*_{\infty}$ will be associated with a unique plasma-coil separation, which can be found using root-finding. Second, on a winding surface, a higher current density results in lower coil-coil separation, which is a common measure of coil complexity.\cite{Paul_2018} To understand why, consider $N$ coils which approximate the current sheet on a winding surface. Via Lagrange's identity, $\| K \| = \| \nabla \Phi \|$. Let us assume the total poloidal current $I_{\mathrm{pol}}$ is distributed uniformly among the coils, so the current in each coil is $I_{\mathrm{pol}}/N$, where $N$ is the number of coils. In the limit of a large number of coils, the current density can be expressed in terms of the distance between coils $d_{cc}$:
\begin{equation}
    \|K\| \equiv  \lim_{N\to\infty} \frac{I_{\mathrm{pol}}}{N d_{cc} } .    
\end{equation}
 In the case of a finite number of coils, this formula remains a good approximation when $d_{cc}$ is small. Therefore,
 \begin{equation}
    K_{\mathrm{max}} \equiv  \|K\|_\infty \approx \frac{I_\mathrm{pol}}{N d_\mathrm{min} },
    \label{eq:Kmax}
 \end{equation}
  where $d_\mathrm{min}$ is the minimum coil-coil distance. 
  To choose $\|K\|^*_{\infty}$, we considered values that have been proposed in detailed engineering studies.
  For example, in ARIES-CS, there are 18 coils (3 unique coils per half field period) and the minimum coil-coil separation is 77.32 cm.\cite{ARIES-CS_report} Using equation (\ref{eq:Kmax}), this separation is equivalent to a maximum current density of 17.16 MA/m. Therefore, we vary $L$ so that $\|K\|_\infty$ = 17.16 MA/m.

\subsection{Additional Methods for finite $\beta$ Configurations}

Before we can calculate \lgradb~and the optimized plasma-coil separation $L_{\mathrm{\texttt{REGCOIL}}}$, we need to calculate two additional terms when $\beta$, the ratio of plasma to magnetic pressure, is nonzero. 
This complication arises because in this case the magnetic field is not generated only by currents in the external coils, and
 the current from the plasma may complicate the relationship between \lgradb~and the plasma-coil separation. Therefore, for this paper we will be careful to decompose the total magnetic field $\mathbf{B}$ into the magnetic field from the coils $\mathbf{B}_{\mathrm{coils}}$, and the magnetic field from the plasma, $\mathbf{B}_{\mathrm{\mathrm{plasma}}}$:
\begin{equation}
   \mathbf{B} = \mathbf{B}_{\mathrm{\mathrm{plasma}}}  + \mathbf{B}_{\mathrm{coils}}.
   \label{eq:sum}
\end{equation}

 To compute \lgradb, we use the virtual casing method  to calculate $\mathbf{B}_{\mathrm{coils}}$ on the LCFS. \cite{shafranov1972use,Virtual_Casing} To summarize the virtual casing approach,  $\mathbf{B}_{\mathrm{coils}}$ is computed  using a magnetic method of images to match the boundary conditions on the plasma surface. As a result, we can find a solution for $\mathbf{B}_{\mathrm{coils}}$ on the last closed flux surface $\delta \Omega$ :

\begin{equation}
     \mathbf{B} _{\mathrm{coils}} (\mathbf{x}) = - \frac{1}{4 \pi} \oint_{\delta \Omega} d^2 p \frac{(\mathbf{n}\times \mathbf{B(p)}) \times (\mathbf{x-p})}{|\mathbf{x-p}|^3},
\end{equation}
 where $\mathbf{p}$ is a point on the LCFS. We used the package \texttt{VIRTUAL\_CASING}\cite{Malhotra_2020, virtualcasing} via \texttt{SIMSOPT}\cite{Landreman2021, SIMSOPT} in order to calculate $\mathbf{B}_{\mathrm{coils}}$ and $\nabla\mathbf{B}_{\mathrm{coils}}$, which we then used to calculate \lgradb~via equation (\ref{eq:lgradb}). 
 
 A similar issue is present in optimizing the external coils using \texttt{REGCOIL}. \texttt{REGCOIL}'s objective function in equation (\ref{eq:objective}) uses the total magnetic field $\mathbf{B}$ as in equation (\ref{eq:sum}). Therefore, we must find $\mathbf{B}_{\mathrm{plasma}}$. \texttt{REGCOIL} is interfaced to the \texttt{BNORM} code from the \texttt{STELLOPT} suite,\cite{STELLOPT} which calculates $\mathbf{B}_{\mathrm{plasma}}$ using data from a \texttt{VMEC} file.\cite{hirshman1983steepest} Then $\mathbf{B}_{\mathrm{coils}}$ is added to  $\mathbf{B}_{\mathrm{plasma}}$ to calculate the objective function \ref{eq:objective} using equation (\ref{eq:sum}). While it would also be possible to use \texttt{VIRTUAL\_CASING} to calculate $\mathbf{B}_{\mathrm{plasma}}$ for \texttt{REGCOIL}, doing so would require an additional software interface, so using \texttt{BNORM} in \texttt{REGCOIL} was more convenient.

\begin{figure}[!h]

    \begin{tikzpicture}[node distance=20mm]
    
    \node (start) [start] {Initialize $\lambda$ and $L$
    (Use \texttt{BNORM} if $\beta > 0$) };
    \node (regcoil) [io, below of=start] {Construct winding surface with separation $L$};
    \node (pro1) [process, below of=regcoil] {Calculate $\Phi$ on winding surface};
    \node (dec1) [decision, below of=pro1, yshift=-0.5pt] { $B_{\mathrm{RMS}}$ = $B_\mathrm{{RMS}}^*$ ?};
    
    \node (pro2a) [decision, below of=dec1, yshift=-0.5cm] {$\|K\|_\infty$ = $\|K\|_\infty^*$ ?};
    
    \node (pro2b) [connect, left of=dec1, xshift=-10mm] {Vary $\lambda$};
    
    \node (test) [connect, right of = dec1, xshift=10mm] {Vary $L$};
    
    \node (out1) [io, below of=pro2a] {Output $\equiv L_{\mathrm{\texttt{REGCOIL}}}$};
    
    \node (stop) [stop, right of=out1, xshift = 2cm] {Stop};
    
    \draw [arrow] (start) -- (regcoil);
    \draw [arrow] (regcoil) -- node[anchor=east]{via $\texttt{REGCOIL}$} (pro1);
    \draw [arrow] (pro1) -- (dec1);
    \draw [arrow] (dec1) -- node[anchor=east] {Yes} (pro2a);
    \draw [arrow] (dec1) -- node[anchor=south] {No} (pro2b);
    \draw [arrow] (pro2b) |- (pro1);
    \draw [arrow] (pro2a) -|node[anchor=west] {No} (test);
    \draw [arrow] (test) |- (regcoil);
    \draw [arrow] (pro2a) --node[anchor=east] {Yes} (out1);
    \draw [arrow] (out1) -- (stop);

    \end{tikzpicture}   
    
    \caption{A flowchart depicting the iterative stage II optimization via \texttt{REGCOIL} to achieve the necessary coil constraints.}

    \label{fig:flowchart}

\end{figure}
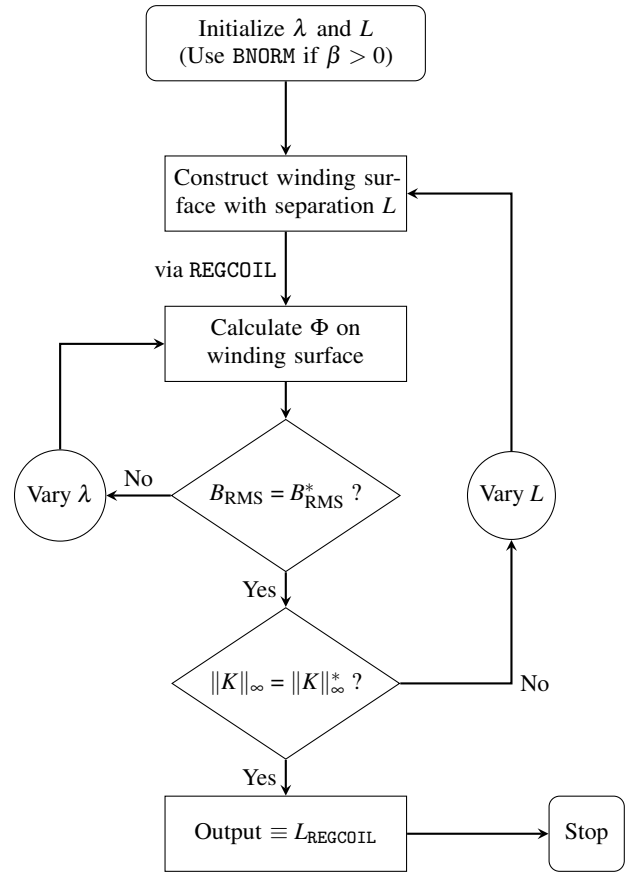

\begin{figure*}
     \centering
     \includegraphics[width=\textwidth]{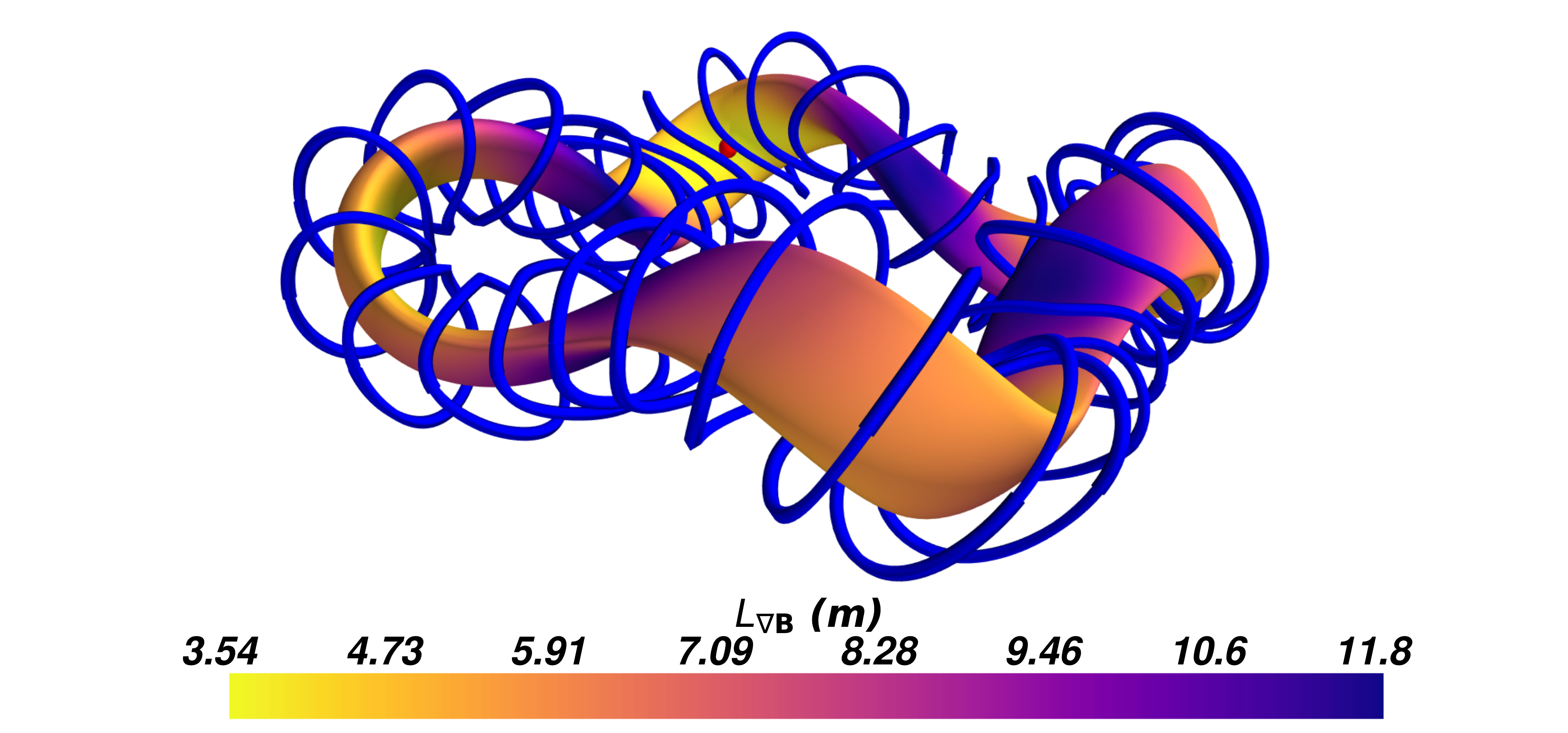}
     \hfill
     \caption{The precise QH plasma configuration\cite{PhysRevLett.128.035001} and the coil arrangements found by \texttt{REGCOIL}. Here $L_\mathrm{\texttt{REGCOIL}}$ is 1.7960 m, as this value was found to be the required plasma-coil separation to satisfy the accuracy and complexity constraints. The red dot is located where \lgradb~is smallest.}
      \label{fig:QHMayavi}
\end{figure*}

\begin{figure*}
     \centering
     \includegraphics[width=0.9\textwidth]{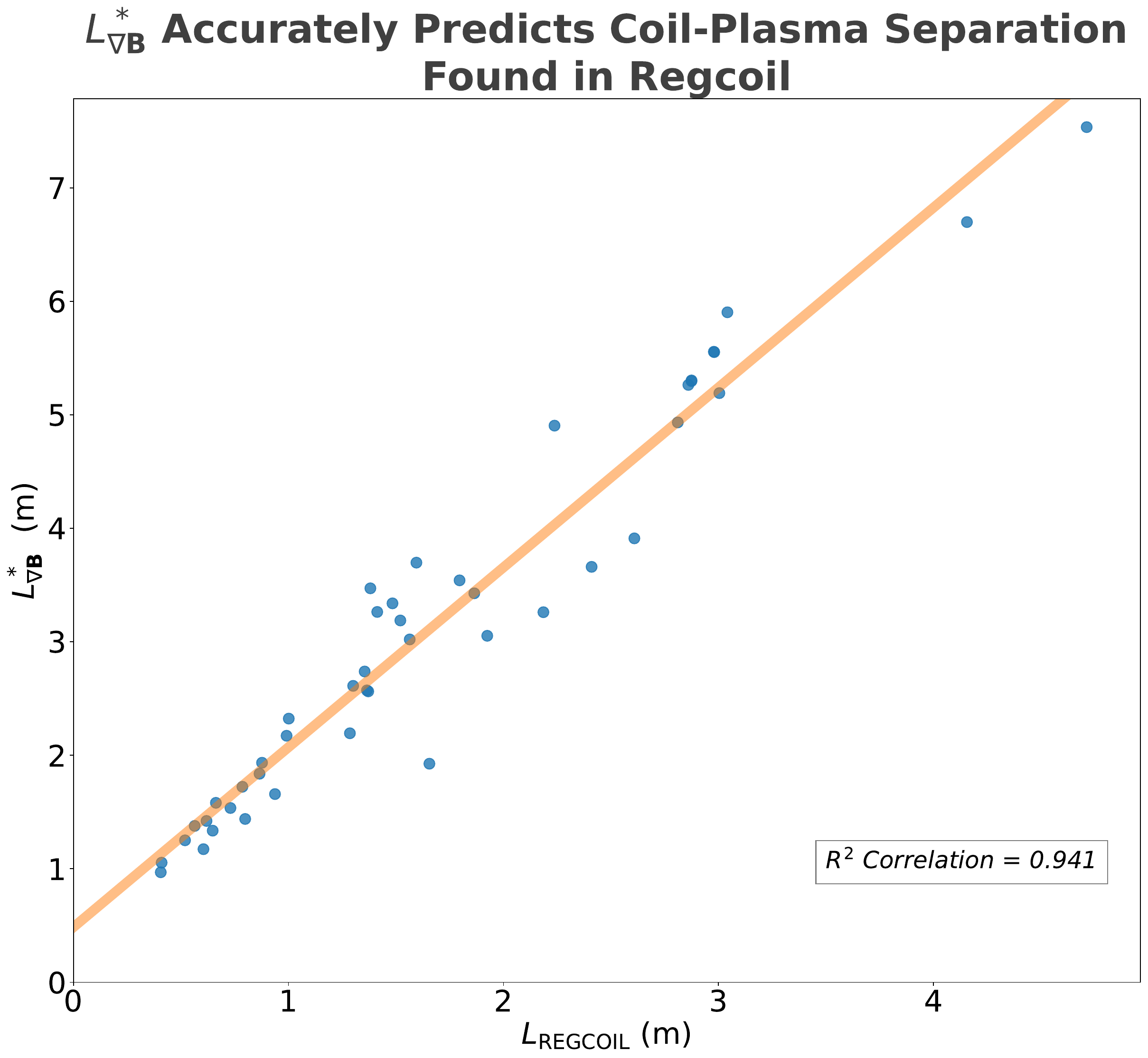}
     \hfill
     \caption{For a wide range of plasma configurations scaled to a minor radius of 1.704 m, we used \texttt{REGCOIL} to perform stage II coil-optimization (subject to constraints in field accuracy and coil complexity). The plasma-coil separation found (labeled $L_\mathrm{\texttt{REGCOIL}}$) has a strong correlation with the smallest value of \lgradb~evaluated on the surface.}
      \label{fig:main plot}
\end{figure*}

\begin{figure*}
     \centering
     \includegraphics[width=\textwidth]{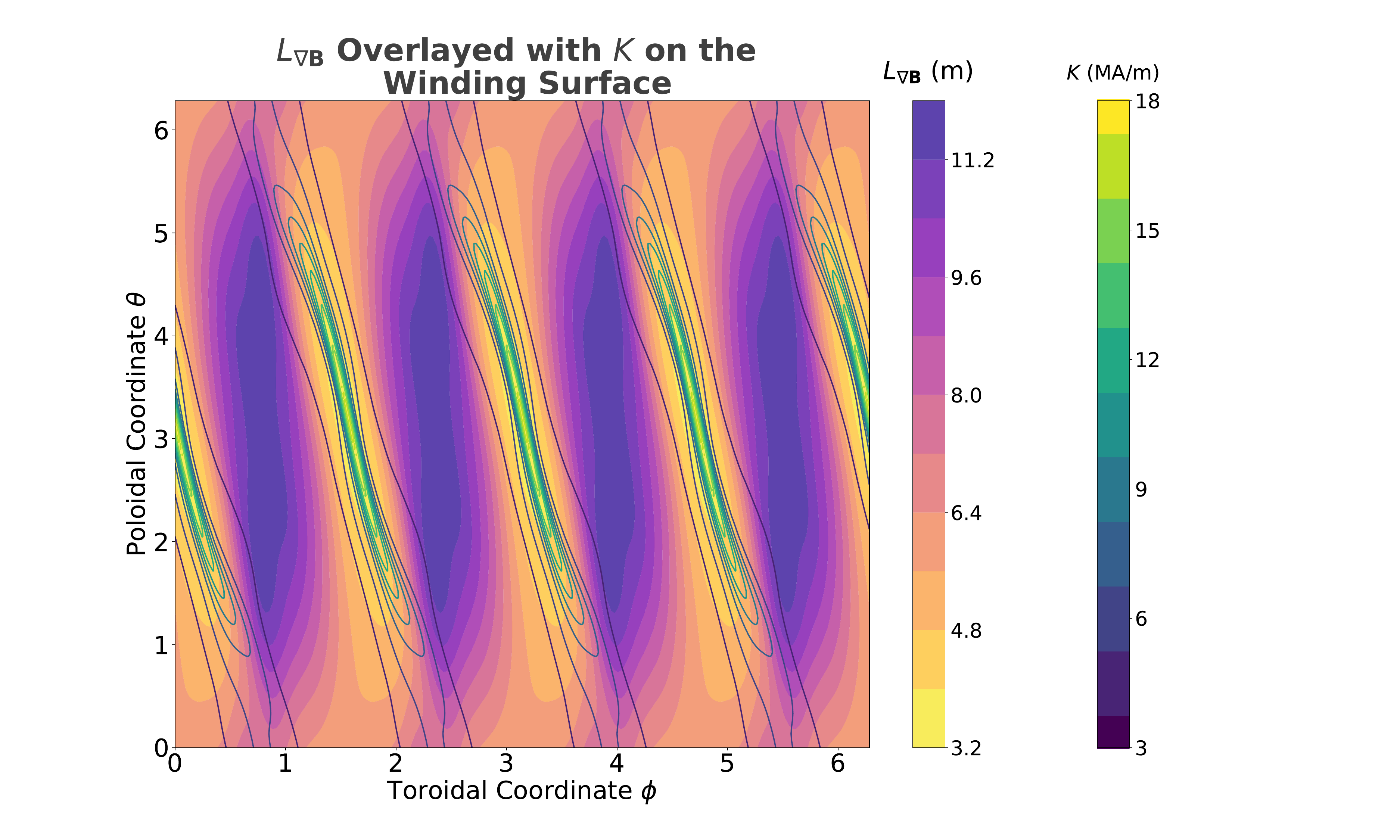}
     \hfill
     \caption{Two contour plots are overlaid for the precise QH stellarator
     and the coil solution of figure \ref{fig:QHMayavi}.
     The filled-in contours indicate \lgradb~on the plasma surface. The unfilled contours  depict the current density $K$ on the winding surface, determined as described in section \ref{section:Methods}. The locations of smallest \lgradb~and largest $K$ coincide, at the inside of the bean cross-section. This spatial correlation is strongest in configurations close to the linear regression line in figure \ref{fig:main plot}.}
      \label{fig:contourK}
\end{figure*}

\subsection{Summary of Optimization Methods}

Given these additional considerations, we now  summarize the process of computing the coil-plasma distance with \texttt{REGCOIL} and comparing the result with  \lgradb. We first choose an initial $\lambda$ and $L$. If $\beta$ $>$ 0, we must also compute $\mathbf{B}_{\mathrm{plasma}}$ using \texttt{BNORM}. We then construct a winding surface offset from the plasma by a uniform distance $L$. We then compute the amplitudes $\Phi_j$ in the Fourier series of the current potential $\Phi$ on the winding surface in order to minimize the loss function $f$ in equation (\ref{eq:objective}). We then perform a search over the regularization parameter $\lambda$ so that the normalized quadratic flux constraint $B_\mathrm{RMS} = B^*_\mathrm{RMS}$ is fulfilled. We found that 0.01 T was sufficient to match the target LCFS, as shown in figure \ref{fig: poincare1}. We then scan over $L$ to fulfill the maximum current density constraint, $\|K\|^*_\infty$ = 17.16 MA/m. We refer to the optimized plasma-coil separation subject to these constraints as $L_{\mathrm{\texttt{REGCOIL}}}$. Refer to figure \ref{fig:flowchart} for a flowchart depicting the optimization process. We checked the numerical resolution needed to ensure convergence of $L_\mathrm{\texttt{RECOIL}}$. We determined that for these parameters, sufficient convergence was achieved when the number of poloidal modes $m_j$ and toroidal modes $n_j$ used to represent the current potential both equal 20, and the number of grid points in each angle necessary to discretize the surface area integrals (referred to within \texttt{REGCOIL}'s code as \texttt{ntheta\_coil}, \texttt{nzeta\_coil}, \texttt{ntheta\_plasma}, and \texttt{nzeta\_plasma}) all equal 96.

We also calculate the magnetic gradient scale length for the plasma configuration. We first calculate $\mathbf{B}_{\mathrm{coils}}$ and $\nabla\mathbf{B}_{\mathrm{coils}}$ using \texttt{VIRTUAL\_CASING} in \texttt{SIMSOPT} if $\beta > 0$. We then calculate \lgradb~on a high resolution grid of points on the LCFS. In Appendix \ref{sec:coordinates}, we describe the method of converting from flux coordinates (like the coordinates used in \texttt{VMEC}\cite{hirshman1983steepest}) to Cartesian coordinates in order to calculate the Frobenius norm. We define
\begin{equation}
    L^*_{\nabla \mathbf{B}} = \min \{L_{\nabla \mathbf{B}}(\mathbf{x}) |\ \mathbf{x} \in \delta \Omega\}
\end{equation}
as the smallest value of \lgradb~on the LCFS, $\delta \Omega$.

We show an example of the optimized coil arrangement and \lgradb~for one configuration, the precise QH stellarator, \cite{PhysRevLett.128.035001} in figure \ref{fig:QHMayavi}.

\section{Results and Discussion}
\label{section:Results}

We have gathered a database of over 40 plasma configurations. These configurations are listed in Appendix \ref{sec:table}. The database includes quasi-axisymmetric stellarators, quasi-helical stellarators, quasi-isodynamic stellarators, tokamaks, and other miscellaneous stellarators. The \texttt{VMEC} files of these configurations and supplemental data have been made available on Zenodo.\cite{Zenodo} For a fair comparison, all configurations have been scaled to the minor radius of the reactor-scale ARIES-CS configuration, $a$ = 1.704 m.\cite{ARIES-CS_report}
(If the configurations are all scaled to the same major radius instead of minor radius, the results that follow are not qualitatively different.)
The volume averaged magnetic field of each configuration in our database is also scaled to the volume averaged magnetic field of ARIES-CS, 5.865 T.
Interestingly, the calculated $L_{\mathrm{\texttt{REGCOIL}}}$  varies by  a factor of 10 across the database. Evidently, there are significant differences among configurations in the feasibility of attaining large coil-plasma separation. 

Figure \ref{fig:main plot} compares the minimum \lgradb~for each configuration with the plasma-coil separation obtained via \texttt{REGCOIL}.
A strong correlation between these two quantities is evident; the correlation coefficient $\mathrm{R}^2 = 0.941$. It is noteworthy that the correlation of figure \ref{fig:main plot} is insensitive to the target $\|K\|^*_\infty$ and  $B^*_{\mathrm{RMS}}$ we choose. Changing either of these constraints within a plausible range changes only the slope and intercept of the linear regression between \lgradb~and $L_{\mathrm{\texttt{REGCOIL}}}$, but has little effect on the strength of the correlation. This correlation is also insensitive to the exact formulation of the magnetic gradient scale length. In section \ref{section:Construction}, we considered some alternative scale lengths to \lgradb. In Appendix \ref{sec:scalelength}, we present similar figures to figure \ref{fig:main plot} calculated with these alternate scale lengths. These scale lengths have slightly weaker correlation with $L_{\texttt{REGCOIL}}$ compared to \lgradb, but are generally still well correlated.

As can be seen in Appendix \ref{sec:table}, there is a trend that the largest coil-plasma distances are possible only with a small number of field periods. Indeed, the two configurations with the largest plasma-coil separation have a single field period, and the next seven configurations with the largest plasma-coil separation all have two field periods. 

Another noteworthy result is that the optimized current density $K$ on the winding surface and \lgradb~calculated on the LCFS are often correlated spatially. For example, in figure \ref{fig:contourK} we have presented overlapping contour plots for $K$ and \lgradb~for one configuration. The two angles on the winding surface $(\theta ', \phi ')$  are defined so that for a position vector $\mathbf{r}'$ on the winding surface,
\begin{equation}
    \theta(\mathbf{r}') , \phi(\mathbf{r}') = \theta, \phi \;\;\; \mathrm{s. t.} \;\;\;   \mathbf{r}(\theta,\phi) + L_\mathrm{\texttt{REGCOIL}} \mathbf{n} = \mathbf{r}',
\end{equation}
where $\mathbf{r}$ is the position vector on the LCFS at the associated point $(\theta,\phi)$, and $\mathbf{n}$ is the outward unit normal vector on the LCFS. For many configurations, the position of maximum current density $\|K\|_\infty$ and the position of the smallest scale length  $L^*_{\nabla \mathbf{B}}$ coincide. Spatial correlation of the contours is strongest near the limiting point (largest $\|K\|_\infty$ and smallest \lgradb),  weaker in regions of lower $\|K\|$ and larger \lgradb. There is usually weaker spatial correlation in configurations that are relatively far from the linear regression line in figure \ref{fig:main plot}. We also found a strong spatial correlation between the location of  $\|K\|_\infty$ and the position of highest radius of curvature, equation (\ref{eq: radius of curvature}). The points of largest $\|K\|_\infty$, smallest \lgradb, and smallest radius of curvature are usually in the inside curve of the bean-shaped cross-section, as shown in figure \ref{fig:QHMayavi}. This relationship matches conventional wisdom that coil engineering constraints are hardest to satisfy in this region.

We should note that the shift caused by accounting for non-vacuum magnetic fields was relatively small. If we do not correct for the difference between $\mathbf{B}$ and $\mathbf{B_\mathrm{coils}}$, both \lgradb~and $L_\mathrm{\texttt{REGCOIL}}$ only shift slightly. This shift is small because the database we used only contained configurations with $\beta \le$  5\%. Therefore,  $\mathbf{B_\mathrm{plasma}}$ was an order of magnitude less than $\mathbf{B_\mathrm{coils}}$. There may be a more significant shift in configurations with higher $ \beta $.

As figure \ref{fig:main plot} shows, a number of configurations have some deviation from the linear regression. Some of these configurations have properties that suggest why \lgradb~or $L_\mathrm{\texttt{REGCOIL}}$ for these configurations may not lie on the best-fit line. 
For example, one point that is somewhat below the best-fit line is the ITER tokamak\cite{10.1063/1.3575626}. This configuration may lie off the trend line because \texttt{REGCOIL} cannot match the target $B_{\mathrm{RMS}}$, which vanishes for axisymmetric configurations. 

Another property that is present among some outliers is coil ripple. For example, the point furthest above the best-fit line is an HSX configuration \cite{doi:10.13182/FST95-A11947086} that includes ripple from its 48 modular coils. We compared this standard HSX configuration to an HSX variant in which the ripple was removed. The shape of HSX's last closed flux surface is defined by a Fourier series,
\begin{align}
  R(\theta, \phi) &= \sum_{m,n} R_{m,n} \cos (m \theta - n_{fp} n \phi),\\
  z(\theta, \phi) &= \sum_{m,n} z_{m,n} \sin (m \theta - n_{fp} n \phi),
\end{align}
where $m$ and $n$ are the poloidal and toroidal mode numbers, respectively, and $n_{fp}$ is the number of field periods. To create the HSX variant, we truncate any boundary mode with $n>4$, which smooths out the field. For this HSX variant, the stage II optimized coils can be placed farther away, and the ripple-less HSX configuration better matches the linear regression. This may point to a limitation in \lgradb.  The magnetic field of HSX only differs from ripple-less HSX on small spatial scales, while \lgradb~primarily captures large spatial scales. Since additional gradients magnify small spatial scales, it may be worthwhile to consider alternative scale lengths based on $\nabla \nabla \mathbf{B}$. 

A final group of outlier configurations in figure \ref{fig:main plot} is associated with relatively large errors in the \texttt{VMEC} solution. Because \lgradb~is calculated using data from \texttt{VMEC}, the \texttt{VMEC} solution must be accurate. We can check the accuracy of these solutions by measuring  $\mathrm{Tr}(\nabla \mathbf{B})$, which should be 0 everywhere, or $\nabla \mathbf{B} - (\nabla \mathbf{B})^ T$, which should be 0 in a vacuum. 
These quantities differ from zero significantly in some of the configurations that are outliers in figure \ref{fig:main plot}.
For example, the CTH configuration with high rotational transform,\cite{peterson2007initial} which is the configuration that is farthest from the line of best fit, has a  $\| \nabla \mathbf{B} - (\nabla \mathbf{B})^ T \|_F =0.19 \|\nabla \mathbf{B}\|_F$. For configurations such as this one with large $\| \nabla \mathbf{B} - (\nabla \mathbf{B})^ T \|_F$, we were unable to obtain satisfactory \texttt{VMEC} solutions at higher resolution, since at higher resolution it is hard to achieve small values of the MHD force residual. This issue is likely specific to \texttt{VMEC}, so these configurations might lie closer to the best-fit line in figure \ref{fig:main plot} if another equilibrium code was used instead.

\section{Summary and Future Work}
\label{Sec:Summary}

Over the course of this paper, we explored the magnetic gradient scale length and its connection to plasma-coil separation. Since a magnetic field decays with distance from the source, the source cannot be much farther than this scale length. The distance to a source is exactly the \lgradb~scale length in the case of an infinite straight wire.

We calculated the minimum magnetic gradient scale length for a wide variety of over 40 plasma configurations, and found it to correlate strongly with the coil-plasma separation, for fixed magnetic field accuracy and coil complexity. For a given configuration, the point of minimum magnetic scale length is typically located on the inside of the bean-shaped cross-section of the plasma, where coil complexity is typically greatest. 
These observations support the use of \lgradb~as a measure of the intrinsic difficulty of producing a given magnetic field with distant coils.
We can now understand why it is difficult to find stage-II coil solutions far from the plasma for certain plasma configurations:
if the configuration has a small \lgradb, it will be hard to find distant coils no matter what specific coil design code is used.

We observe a general inverse relationship between the number of field periods and the plasma-coil separation.
In particular, the nine configurations with largest coil-to-plasma distance compared to the minor radius all have only one or two field periods. 
This makes one- or two-field-period configurations attractive for reactors, where a blanket and neutron shielding are required between the coils and plasma.

We plan on further studying the nature of the magnetic gradient scale length \lgradb. There is also merit in studying the norm of $\nabla \nabla \mathbf{B}$, which is a rank 3 tensor. However, $\nabla \nabla \mathbf{B}$  may be more susceptible to the noise in the magnetic field due to discretization error in the MHD equilibrium solution. In the future, we would like to study how the slope of the linear regression of figure \ref{fig:main plot} changes as $B^*_\mathrm{RMS}$ and $\|K\|_\infty^*$ change. 

In addition, it would be natural to try including \lgradb~in the objective functions of Stage I optimizations (i.e., optimizing the plasma shape without explicitly considering coil shapes), with the goal of increasing the plasma-coil separation. Compared to \texttt{REGCOIL} or other coil design codes, using \lgradb~to approximate plasma-coil separation requires few assumptions, since we do not have to set constraints on the quadratic flux accuracy or the highest current density. In addition to \lgradb~taking less time to calculate than plasma-coil separation, it is easier to differentiate through \lgradb, which would further shorten optimization time.

\section*{Acknowledgements}

We are grateful to the many people who contributed MHD configurations to this study.
We would also like to thank Stefan Buller and Byoungchan Jang for comments on the manuscript.
This work was supported by the U.S. Department of Energy, Office of Science, Office of Fusion Energy Science, under award number DE-FG02-93ER54197.
This research used resources of the National Energy Research Scientific Computing Center (NERSC), a U.S. Department of Energy Office of Science User Facility located at Lawrence Berkeley National Laboratory, operated under Contract No. DE-AC02-05CH11231 using NERSC award FES-ERCAP-mp217-2023.

\appendix

\section{Magnetic Field of a Circular Wire }
\label{sec:circular}

 Consider a circular wire of radius $a$ in the $x$-$y$ plane. The magnetic field is as follows:\cite{article}
\begin{gather}
    \rho^2 = x^2 +y^2 ;\  r^2 = x^2 +y^2 +z^2 ; \ \alpha^2 = a^2 +r^2 +z^2 - 2 a \rho, \\
    \beta^2 = a^2 +r^2 + 2 a \rho ;\  k^2 = 1 - \frac{\alpha^2} {\beta^2}, \\
    B_x =\frac{\mu_0 I}{\pi} \frac{x z}{2 \alpha^2 \beta \rho^2}\biggl( (a^2 +r^2)E(k^2) - \alpha^2 K(k^2) \biggr), \\
    B_y = \frac{\mu_0 I}{\pi} \frac{y z}{2 \alpha^2 \beta \rho^2}\biggl( (a^2 +r^2)E(k^2) - \alpha^2 K(k^2) \biggr), \\
    B_z = \frac{\mu_0 I}{\pi} \frac{1}{2\alpha^2 \beta} \biggl( (a^2 - r^2) E(k^2) + \alpha^2 K(k^2)\biggr),
\end{gather}
where $K(k^2)$ and $E(k^2)$ are the complete elliptic integrals of the first and second kind, respectively. At the center of the circular coil, \lgradb $\rightarrow \infty$. The region in the neighborhood surrounding the coil has \lgradb~values that are approximately the distance to the coil, since the circular wire can locally be approximated by a straight wire in the limit that the distance from the coil is small compared to its radius.
\begin{figure}
     \centering
         \centering
         \includegraphics[width= 0.5 \textwidth]{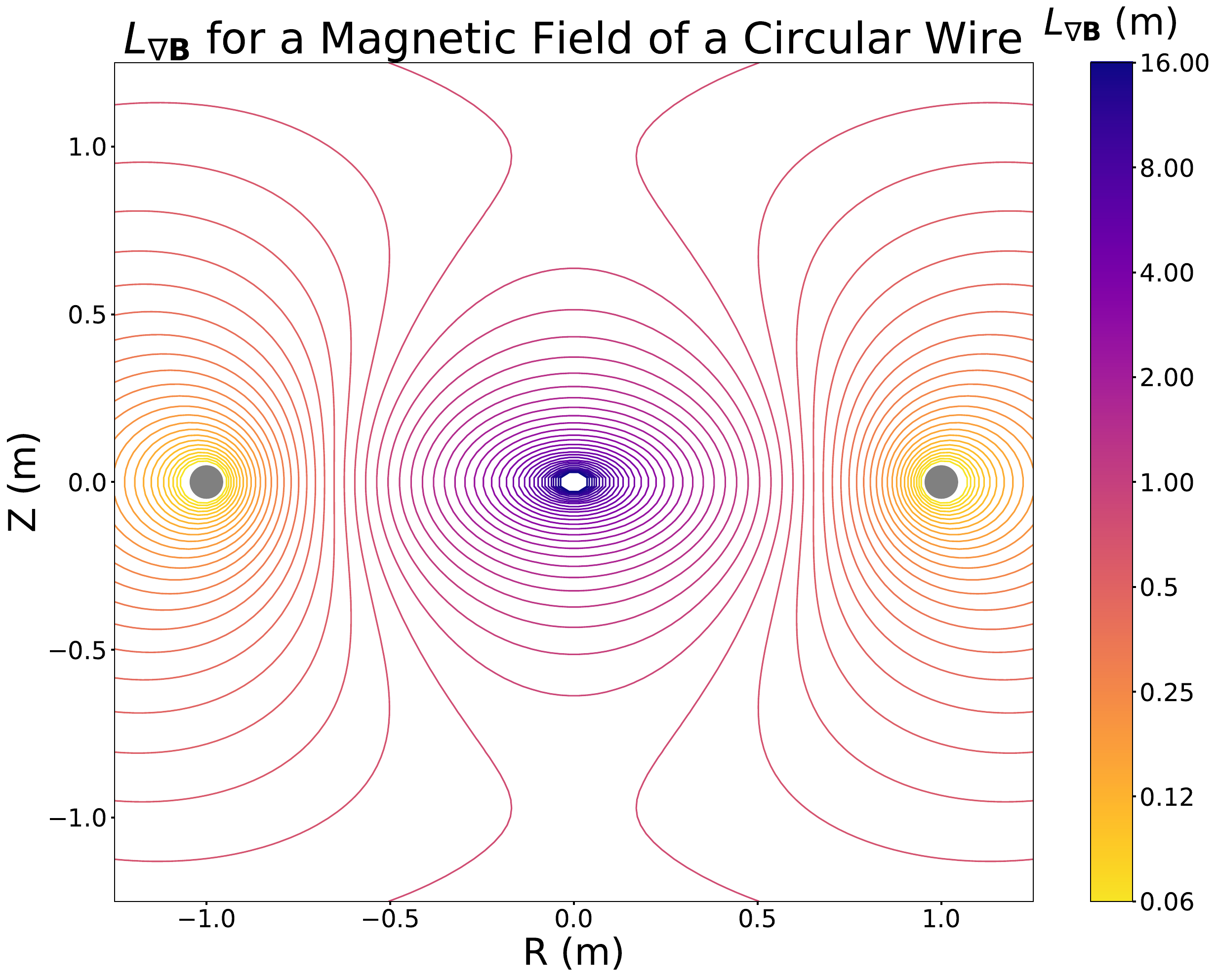}
     \hfill
     \caption{The \lgradb~scale length for the magnetic field generated by a circular wire. The circle has a radius of 1 m and lies in the $x$-$y$ plane. Grey dots signify the location of the wire.}
    \label{fig:circ}
\end{figure}
\begin{figure*}
     \centering
         \centering
         \includegraphics[width= 0.75\textwidth]{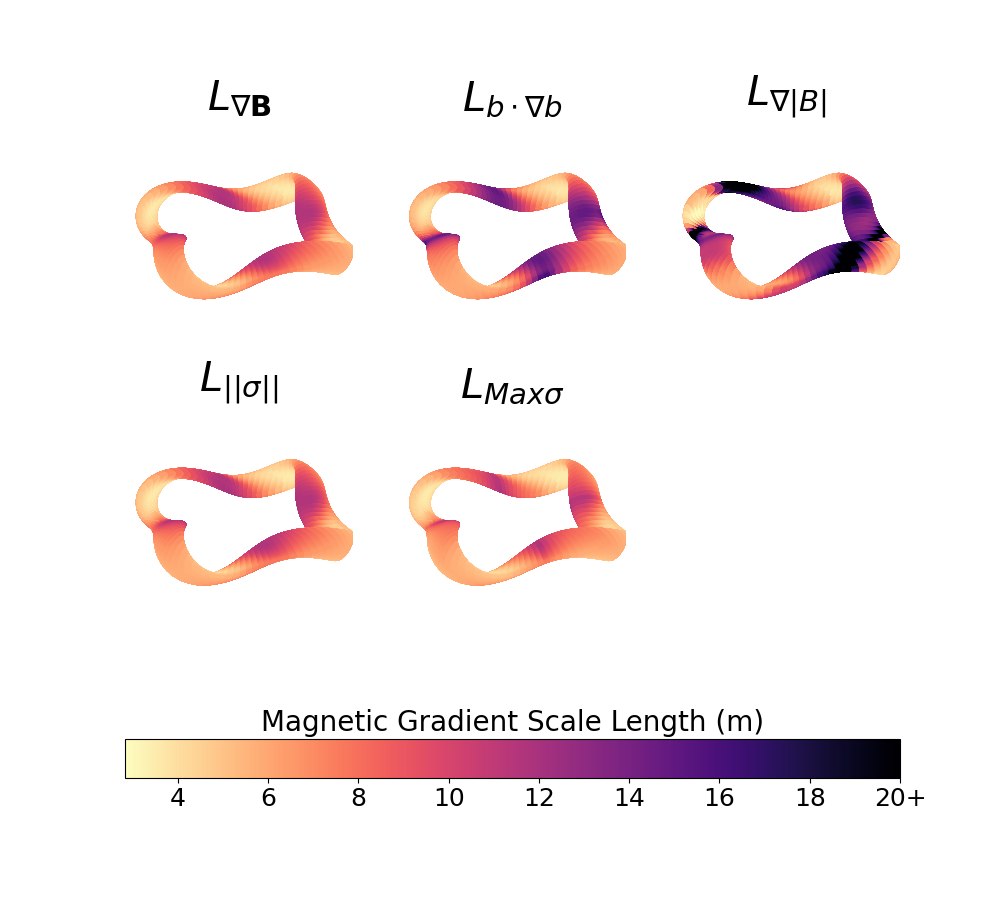}
     \hfill
     \caption{A comparison of magnetic gradient scale lengths for the precise QH plasma configuration.\cite{PhysRevLett.128.035001} For clarity, the colormap is bounded so that any scale length larger than 20 m is truncated to 20 m.}
     \label{fig:scales}
\end{figure*}
In figure \ref{fig:circ}, we have plotted the scale length on the $x$-$z$ plane. As this figure shows, \lgradb~approximates the distance to the coil best when measured close to the coil. 

\section{Alternative Magnetic Gradient Scale Lengths}
\label{sec:scalelength}

\begin{figure*}
     \centering
         \begin{subfigure}{0.33\textwidth}
         \centering
         \includegraphics[width=1.0 \textwidth]{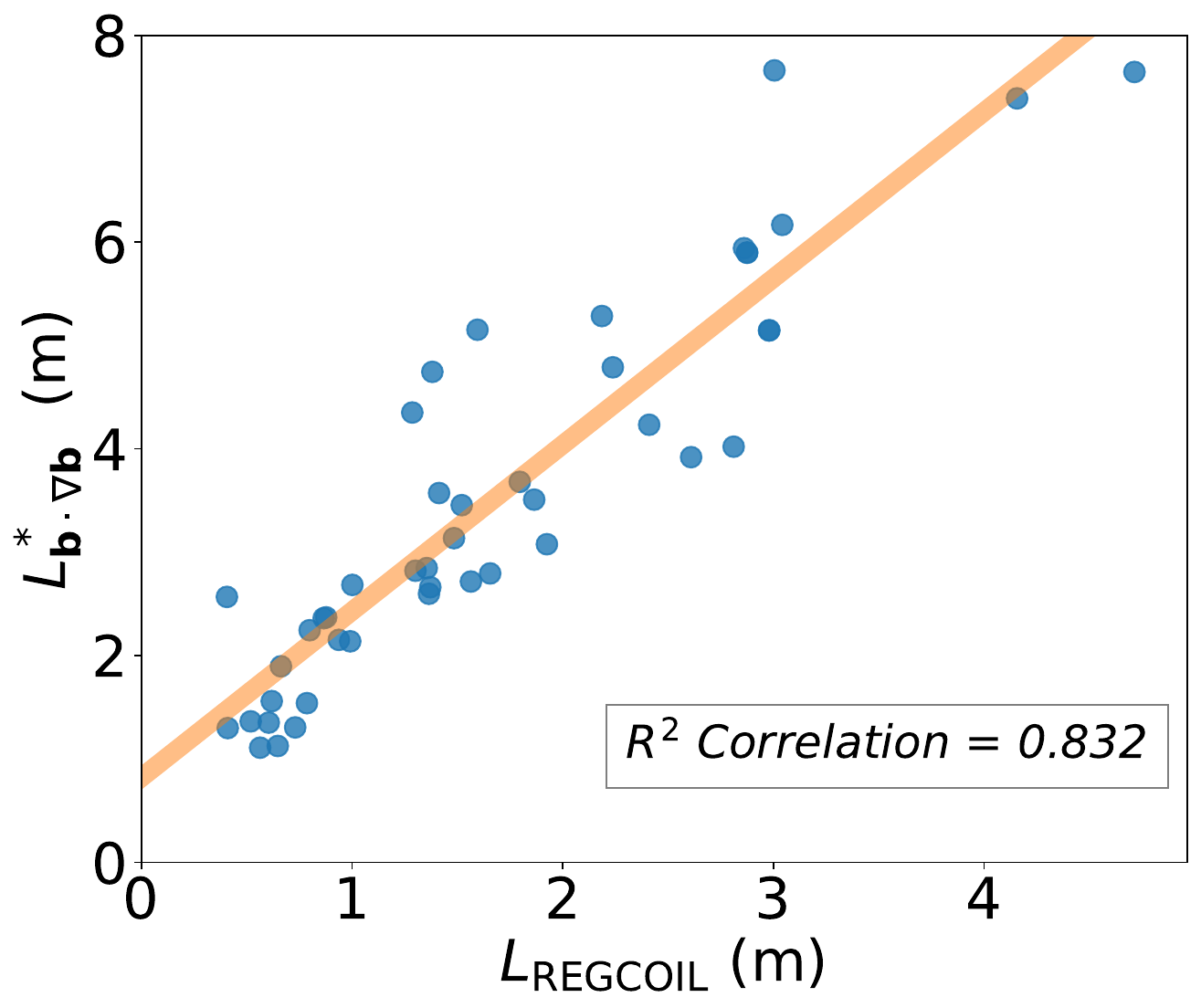}
         \caption{}
         \label{fig: moneyplotcurvature}
     \end{subfigure}
     \begin{subfigure}{0.33 \textwidth}
         \centering
         \includegraphics[width=1.0\textwidth]{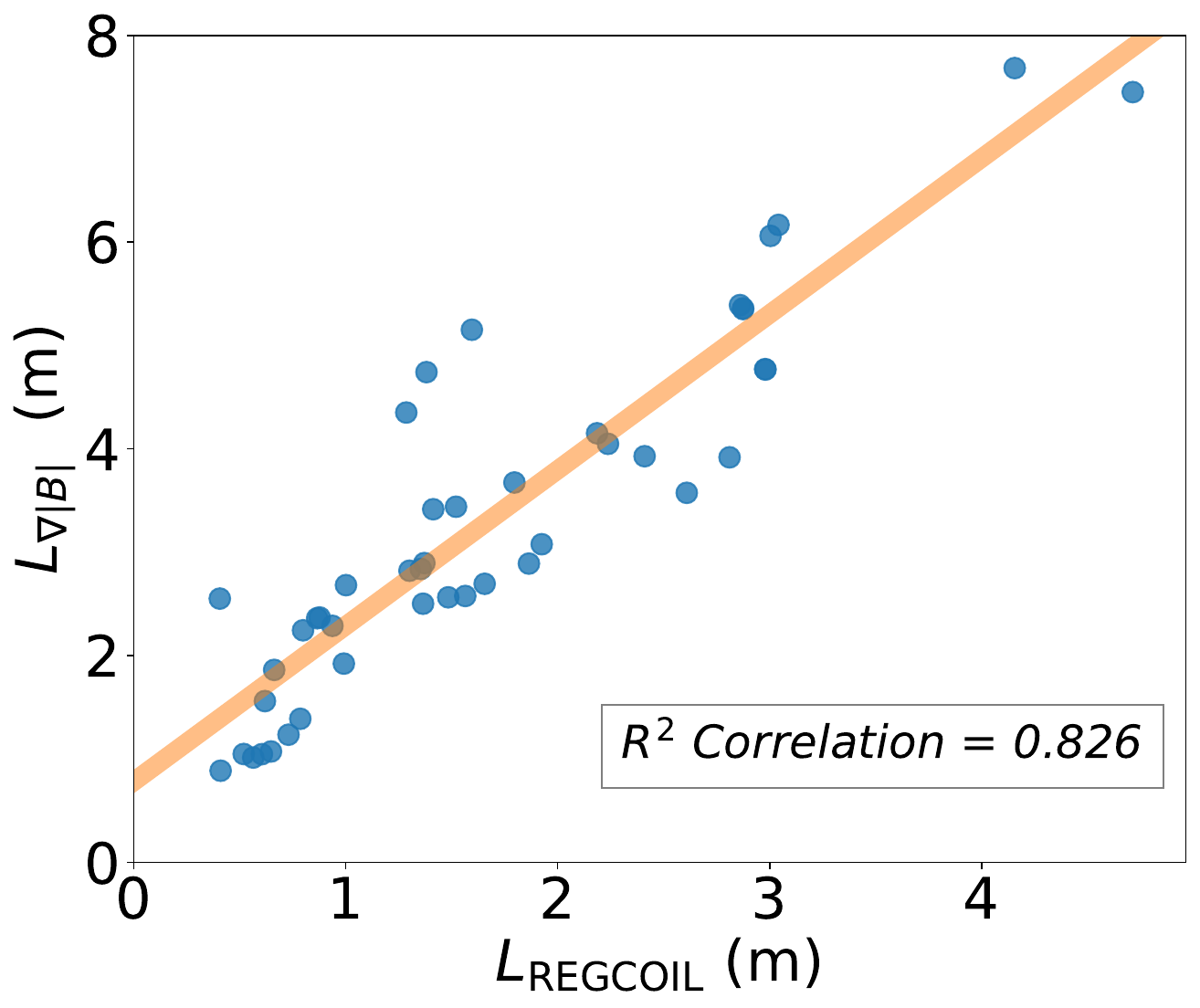}
         \caption{}
         \label{fig: moneyplotscalar}
     \end{subfigure}
     \hfill
    \begin{subfigure}{0.33\textwidth}
     \centering
     \includegraphics[width=1.0\textwidth]{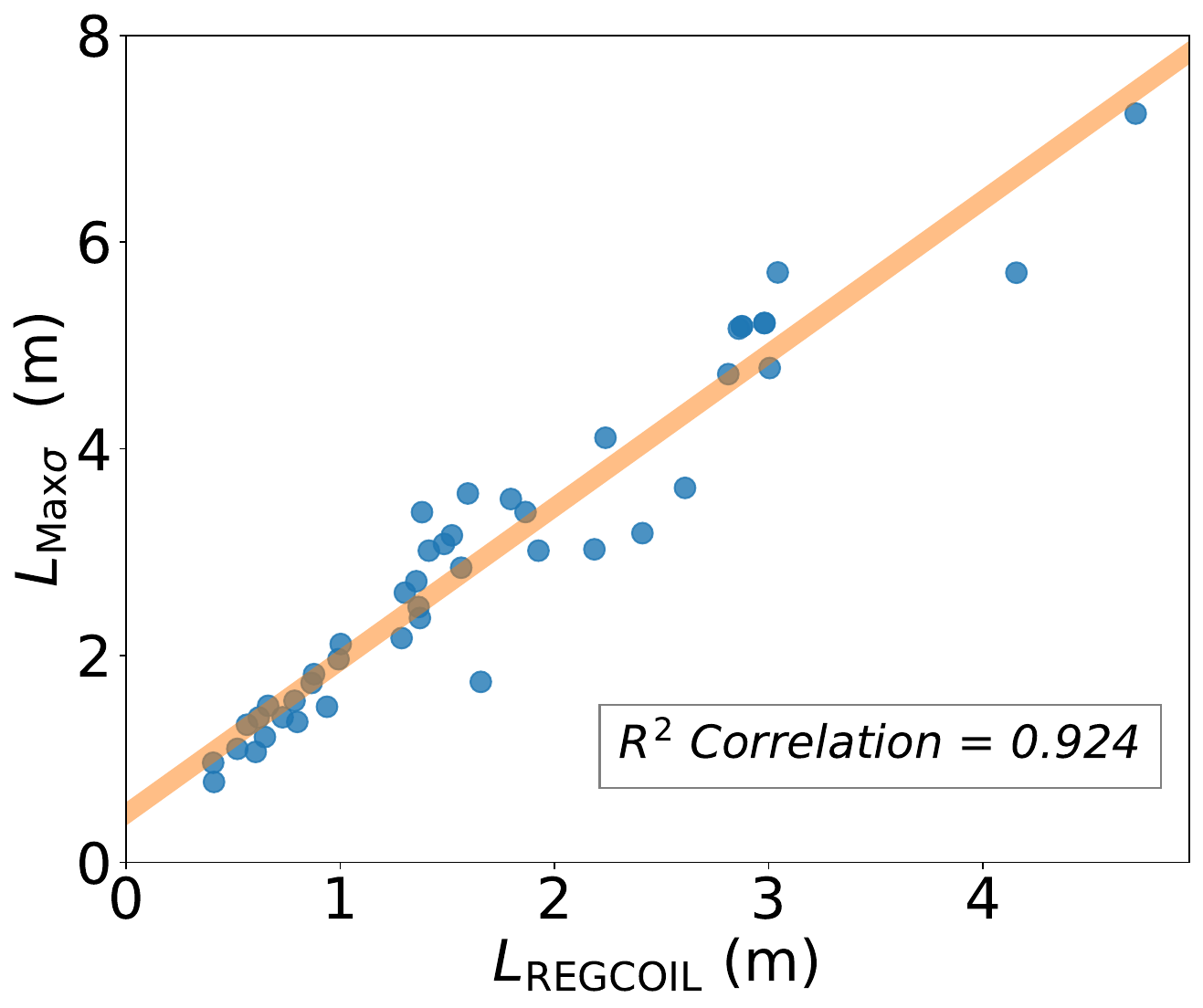}
     \caption{}
     \label{fig: moneyplotsvmax}
    \end{subfigure}
     \hfill
     \caption{For a wide range of plasma configurations scaled to a minor radius of 1.704 m, we used \texttt{REGCOIL} to perform stage II coil-optimization (subject to constraints in field accuracy and coil complexity). The plasma-coil separation (labeled $L_{\texttt{REGCOIL}}$) is correlated with (a) $L_{\mathbf{b} \cdot \nabla \mathbf{b}}$ from equation (\ref{eq:bdotb}), (b) $L_{\nabla |B|}$ from equation (\ref{|B|}), and (c) $ L_{\mathrm{Max}\sigma}$ from equation (\ref{Max}).}
         \label{fig: altmoney}
\end{figure*}

Besides \lgradb, a number of alternative scale lengths that could be considered are
\begin{align}
    \label{eq:bdotb}
    L_{\mathbf{b} \cdot \nabla \mathbf{b}} &= \frac{1}{\|\mathbf{b} \cdot \nabla \mathbf{b}\|},\\
    \label{|B|}
    L_{\nabla |B|} &=  \frac{B}{\|\nabla B\|},\\
    \label{sigma}
    L_{\|\sigma\| }&= \frac{\sqrt{2} B}{ ( \Sigma_i \sigma_i^2)^{1/2}},\\
    \label{Max}
    L_{\mathrm{Max}\sigma} &= \frac{B}{\mathrm{Max}[\sigma_i]},
\end{align}
where $\sigma_i$ are the 3 singular values of the matrix $\nabla \mathbf{B}$. 
Note that $L_{\|\sigma\| }=L_{\nabla\mathbf{B}}$ for a vacuum field.
In figure \ref{fig:scales}, we have plotted these alternate scale lengths on the surface of the precise QH configuration.  All these alternate scale lengths are smallest on the inside of the bean cross-section, which is where \lgradb~was previously mentioned to be smallest in section \ref{section:Results}. At points where \lgradb~is high, only equations \ref{Max} and \ref{sigma} are similar in value to \lgradb. For \ref{|B|} and \ref{eq:bdotb}, the locations of largest scale length are shifted slightly compared to \lgradb.
We find there is a significant correlation between each of these alternate scale lengths and $L_\mathrm{\texttt{REGCOIL}}$, although \lgradb~has the strongest correlation.
In figure \ref{fig: altmoney} we have plotted these alternate scale lengths compared to $L_\mathrm{\texttt{REGCOIL}}$. 

\section{Conversion Between Flux Coordinates and Cartesian Coordinates}
\label{sec:coordinates}

Here we show a method to compute the $\nabla\mathbf{B}$ tensor elements in Cartesian coordinates given $\mathbf{B}$ in flux coordinates $(s, \theta, \phi)$, where $s$ is the flux surface coordinate,  $\theta$ is any poloidal angle coordinate,  and $\phi$ is the traditional azimuthal angle coordinate.\cite{imbertgerard2020introduction} For nested flux surfaces to exist, $\mathbf{B} \cdot \nabla s = 0$. Therefore, the magnetic field can be written in terms of its contravariant components as
\begin{equation}
    \mathbf{B} = B^\theta \frac{\partial \mathbf{r}}{d \theta} +  B^\phi \frac{\partial \mathbf{r}}{d \phi},
\end{equation}
where $\mathbf{r} = R \hat{\mathbf{R}} + z \hat{\mathbf{z}}$ is the position vector, and $(R,\phi,z)$ are cylindrical coordinates. Applying the chain rule to $\mathbf{r}$,
\begin{equation}
   \mathbf{B} = B^\theta \biggl( \frac{\partial R}{d \theta} \hat{\mathbf{R}} + \frac{\partial z}{d \theta} \hat{\mathbf{z}} \biggr) +  B^\phi \biggl(\frac{\partial R}{\partial \phi} \hat{\mathbf{R}} + R \hat{\pmb \phi} + \frac{\partial z}{\partial \phi} \hat{\mathbf{z}} \biggr).
\end{equation}
Therefore,
\begin{align}
B_x &= B^\theta \frac{\partial R}{\partial \theta}\cos \phi + B^\phi \frac{\partial R}{\partial \phi}\cos \phi - B^\phi R \sin \phi,\\
B_y &= B^\theta \frac{\partial R}{\partial \theta}\sin \phi + B^\phi \frac{\partial R}{\partial \phi}\sin \phi + B^\phi R \cos \phi,\\
B_z &= B^\theta \frac{\partial z}{\partial \theta} + B^\phi \frac{\partial z}{\partial \phi},
\end{align}

The representation of $\nabla \mathbf{B}$ as a matrix has 9 components, and the Cartesian form of the matrix is shown in equation (\ref{gradbmatrix}). For simplicity, we shall focus on the formula for the first column of the Cartesian gradient matrix, $\nabla B_x$, but the same steps can be applied to find the other columns of the matrix.

Writing the gradient of $B_x(s, \theta,\phi)$ in terms of flux coordinates, 
\begin{equation}
    \nabla B_x = \frac{\partial B_x}{\partial s} \nabla s + \frac{\partial B_x}{\partial \theta} \nabla \theta  +\frac{\partial B_x}{\partial \phi} \nabla \phi.
    \label{gradb_x}
\end{equation}
Let us find formulas for each term in equation (\ref{gradb_x}) and then convert to cylindrical coordinates. We use the dual relation between $\nabla s$ and the contravariant basis vectors:
\begin{equation}
    \nabla s =\frac{1}{\sqrt{g}} \biggl( \frac{\partial \mathbf{r}}{\partial \theta} \times \frac{\partial \mathbf{r}}{\partial \phi} \biggr),
    \label{eq: nabla}
\end{equation}
where $\sqrt{g} = (\nabla s \times \nabla \theta \cdot \nabla \phi)^{-1}$. The formulas for $\nabla \theta$ and $\nabla \phi$ follow from equation (\ref{eq: nabla}) by cyclic permutation.
We can now convert to cylindrical coordinates:
\begin{equation}
    \nabla s =\frac{1}{\sqrt{g}} \biggl[ \frac{\partial R}{\partial \theta} R \hat{\mathbf{z}} + \biggl(\frac{\partial R}{\partial \phi} \frac{\partial z}{\partial \theta} - \frac{\partial R}{\partial \theta} \frac{\partial z}{\partial \phi}\biggr) \hat{\pmb \phi} - \frac{\partial z}{\partial \theta}R \hat{\mathbf{R}} \biggr].
\end{equation}
Using the same logic,
\begin{equation}
  \nabla \theta =\frac{1}{\sqrt{g}} \biggl[ R \frac{\partial z}{\partial s}  \hat{\mathbf{R}} + \biggl(\frac{\partial R}{\partial s} \frac{\partial z}{\partial \phi} - \frac{\partial R}{\partial \phi} \frac{\partial z}{\partial s}\biggr)\hat{\pmb \phi} - R \frac{\partial R}{\partial s} \hat{\mathbf{z}} \biggr],
\end{equation}
\begin{equation}
  \nabla \phi = \frac{1}{r} \hat{\pmb \phi}.
\end{equation}
We can then use the chain rule to find
\begin{align}
    \begin{split}
     \frac{\partial B_x}{\partial s} =& \frac{\partial B^\theta}{\partial s} \frac{\partial R}{\partial \theta} \cos \phi + B^\theta \frac{\partial^2 R}{\partial s \partial \theta} \cos \phi + 
      \frac{\partial B^\phi}{\partial s} \frac{\partial R}{\partial \phi} \cos \phi \\& + B^\phi \frac{\partial^2 R}{\partial s \partial \phi} \cos \phi - \frac{\partial B^\phi}{\partial s} R \sin \phi - B^\phi \frac{\partial R}{\partial s} \sin \phi ,    
    \end{split} \\
      \begin{split}
         \frac{\partial B_x}{\partial \theta} =&  \frac{\partial B^\theta}{\partial \theta} \frac{\partial R}{\partial \theta} \cos \phi + B^\theta \frac{\partial^2 R}{\partial \theta^2} \cos \phi + 
      \frac{\partial B^\phi}{\partial \theta} \frac{\partial R}{\partial \phi} \cos \phi   \\&+ B^\phi \frac{\partial^2 R}{ \partial \phi \partial \theta} \cos \phi - \frac{\partial B^\phi}{\partial \theta} R \sin \phi - B^\phi \frac{\partial R}{\partial \theta} \sin \phi ,
    \end{split}\\
        \begin{split}
         \frac{\partial B_x}{\partial \phi} =&  \frac{\partial B^\theta}{\partial \phi} \frac{\partial R}{\partial \theta} \cos \phi + B^\theta \frac{\partial^2 R}{\partial \phi \partial \theta} \cos \phi
         - B^\theta \frac{\partial R}{\partial \theta} \sin \phi \\ &
    + \frac{\partial B^\phi}{\partial \phi} \frac{\partial R}{\partial \phi} \cos \phi + B^\phi \frac{\partial^2 R}{ \partial \phi^2} \cos \phi
     - B^\phi \frac{\partial R}{ \partial \phi} \sin \phi \\& - \frac{\partial B^\phi}{\partial \phi} R \sin \phi - B^\phi \frac{\partial R}{\partial \phi} \sin \phi - B^\phi R \cos \phi .
    \end{split}
\end{align}
We can substitute these formulas into equation (\ref{gradb_x}) in order to calculate $\partial B_x/\partial x$, $\partial B_x/\partial y$, and $\partial B_x/\partial z$. We can use similar steps to calculate any of the other six elements of the matrix $\nabla \mathbf{B}$. 
\clearpage
\onecolumngrid
\vspace{25px}
\section{Table of Plasma Configurations}
\label{sec:table}
Below is a table with a brief description of all configurations used in this paper. They are arranged in order of increasing $L_\mathrm{\texttt{{REGCOIL}}}$. Both the number of field periods (NFP) and the smallest magnetic gradient scale length ($L^*_{\nabla \mathbf{B}}$) are recorded.

\begin{table}[!h]
\begin{tabular}{lrrr}
\toprule
\textbf{Description} &
  \multicolumn{1}{l}{\textbf{NFP}} &
  \multicolumn{1}{l}{\textbf{$L_{\texttt{REGCOIL}}$} (m)} &
  \textbf{$L^*_{\nabla \mathbf{B}}$ (m)}  \\* \midrule
\bottomrule
nfp=4 quasi-helical (QH) configuration by Ku \& Boozer \cite{Ku_2011} & 4 & 0.4060 & 0.9691 \\
Unpublished QH configuration from Michael Drevlak & 5 & 0.4099 & 1.0538 \\
Columbia Non-Neutral Torus (CNT)\cite{PhysRevLett.88.205002} & 2 & 0.5189 & 1.2507 \\
Tokamak de la Junta II (TJ-II) \cite{doi:10.13182/FST17-131-139} & 4 & 0.5638 & 1.3777 \\
Wistell-B, Bader et al. \cite{Bader_2021} & 5 & 0.6047 & 1.1726 \\
Quasi-axisymmetric (QA) configuration designed by Paul Garabedian \cite{garabedian2008three} & 2 & 0.6188 & 1.4214 \\
Large Helical Device (LHD), major radius 3.60m \cite{iiyoshi1999overview} & 10 & 0.6475 & 1.3358 \\
Quasi-Poloidal Stellarator (QPS)\cite{1027687} & 2 & 0.6625 & 1.5812 \\
LHD, major radius 3.53m \cite{iiyoshi1999overview} & 10 & 0.7302 & 1.5354 \\
LHD, major radius 3.75m \cite{iiyoshi1999overview} & 10 & 0.7858 & 1.7226 \\
Henneberg et al. QA \cite{Henneberg_2019} & 2 & 0.7987 & 1.4390 \\
Advanced Research Innovation and Evaluation Study-Compact Stellarator (ARIES-CS) \cite{ARIES-CS_report} & 3 & 0.8655 & 1.8375 \\
National Compact Stellarator Experiment (NCSX) stage-1 optimization result (known as LI383)\cite{NELSON2003169} & 3 & 0.8771 & 1.9343 \\
The first quasisymmetric configuration found\cite{NUHRENBERG1988113} & 6 & 0.9374 & 1.6582 \\
Advanced Toroidal Facility (ATF) \cite{doi:10.13182/FST85-A40084} & 12 & 0.9913 & 2.1721 \\
NCSX free-boundary (c09r00) \cite{NELSON2003169} & 3 & 1.0015 & 2.3233 \\
Wendelstein 7-X (W7-X), without coil ripple\cite{Drevlak} & 5 & 1.2858 & 2.1941 \\
Landreman, Buller, \& Drevlak, QH, 5\% beta \cite{10.1063/5.0098166} & 4 & 1.3009 & 2.6120 \\
Landreman, Buller, \& Drevlak, QH, vacuum \cite{10.1063/5.0098166} & 4 & 1.3545 & 2.7385 \\
Boundary constructed by near-axis expansion. Vacuum QH with nfp=4 & 3 & 1.3650 & 2.5722 \\
Goodman et al. Quasi-isodynamic (QI) configuration with nfp=3 \cite{goodman2022constructing} & 4 & 1.3712 & 2.5633 \\
W7-X standard configuration. Vacuum, with coil ripple \cite{doi:10.13182/FST90-A29178} & 5 & 1.3811 & 3.4715 \\
Quasi-Isodynamic (QI) configuration from CIEMAT \cite{Sánchez_2023} & 4 & 1.4130 & 3.2634 \\
Chinese First Quasiaxisymmetric Stellarator (CQFS) \cite{CFQS_coils} & 2 & 1.4839 & 3.3392 \\
Landreman \& Paul, QH with magnetic well\cite{PhysRevLett.128.035001} & 4 & 1.5206 & 3.1882 \\
Wistell-A. Bader et al. \cite{BaderQH} & 4 & 1.5641 & 3.0210 \\
W7-X "high narrow mirror" configuration\cite{Drevlak_2014} & 5 & 1.5952 & 3.6979 \\
Compact Toroidal Hybrid (CTH) Stellarator, vacuum, with low rotational transform \cite{peterson2007initial} & 5 & 1.6556 & 1.9259 \\
Landreman \& Paul, precise QH \cite{PhysRevLett.128.035001} & 4 & 1.7960 & 3.5418 \\
Unpublished nfp=3 QH & 3 & 1.8644 & 3.4277 \\
Up-down-symmetric ITER-like configuration\cite{10.1063/1.3575626} & 1 & 1.9248 & 3.0531 \\
Evolutive Stellarator of Lorraine (ESTELL) \cite{https://doi.org/10.1002/ctpp.201200055} & 2 & 2.1860 & 3.2610 \\
Helically Symmetric Experiment (HSX), standard configuration, vacuum, with coil ripple\cite{doi:10.13182/FST95-A11947086} & 4 & 2.2377 & 4.9052 \\
Compact Toroidal Hybrid (CTH) stellarator, vacuum, with high rotational transform \cite{peterson2007initial} & 5 & 2.4102 & 3.6607 \\
Boundary constructed by near-axis expansion. Vacuum QH with nfp=3\cite{landreman_2022} & 3 & 2.6091 & 3.9118 \\
HSX, standard configuration, vacuum, without coil ripple & 4 & 2.8111 & 4.9336 \\
Vacuum QA configuration with 16 coils from Giuliani et al. Coil length 24m. \cite{giuliani_wechsung_stadler_cerfon_landreman_2022} & 2 & 2.8602 & 5.2643 \\
Landreman \& Paul, precise QA \cite{PhysRevLett.128.035001} & 2 & 2.8748 & 5.2977 \\
Wechsung et al. QA without magnetic well, coil length 24m. \cite{doi:10.1073/pnas.2202084119} & 2 & 2.8750 & 5.3037 \\
Wechsung et al. QA with magnetic well, coil length 24m \cite{doi:10.1073/pnas.2202084119} & 2 & 2.9790 & 5.5563 \\
Landreman \& Paul QA with magnetic well.\cite{PhysRevLett.128.035001} & 2 & 2.9806 & 5.5532 \\
Goodman et al. Quasi-isodynamic configuration with nfp=2 \cite{goodman2022constructing} & 2 & 3.0045 & 5.1919 \\
Landreman, Buller \& Drevlak, QA, 2.5\% beta \cite{10.1063/5.0098166} & 2 & 3.0419 & 5.9042 \\
Goodman et al. Quasi-isodynamic configuration with nfp=1 \cite{goodman2022constructing} & 1 & 4.1563 & 6.6993 \\
Jorge et al. Quasi-isodynamic configuration with nfp=1 \cite{JorgeQI} & 1 & 4.7133 & 7.5360
\end{tabular}
\end{table}
\twocolumngrid
\section*{References}
\nocite{*}
\bibliography{MagneticGradient}
\end{document}